\documentclass[11pt, a4paper]{article}
\usepackage[margin=25mm]{geometry}
\usepackage{lmodern}
\usepackage{amsthm}
\usepackage{amsmath}
\usepackage{amsfonts}
\usepackage[shortlabels]{enumitem}
\usepackage{amssymb}%
\usepackage{mathrsfs}
\usepackage{textcomp}
\usepackage{authblk}
\usepackage{graphicx}
\usepackage[absolute]{textpos}
\usepackage{xcolor}
\usepackage{soul}
\usepackage[numbers,sort&compress]{natbib}
\usepackage{mathtools}
\usepackage[colorlinks=true, linkcolor=blue, urlcolor=blue, citecolor=blue]{hyperref}
\usepackage{bm}
\usepackage{bbm}
\usepackage{epstopdf}

\newcounter{subeq}
\renewcommand{\thesubeq}{\theequation\alph{subeq}}
\newcommand{\newsubeqblock}{\setcounter{subeq}{0}\refstepcounter{equation}}
\newcommand{\mysubeq}{\refstepcounter{subeq}\tag{\thesubeq}}

\title{{\color{black}Exploring external rarefied gas flows through the method of fundamental solutions}}

\author[1]{Himanshi\thanks{\href{phd2001141003@iiti.ac.in}{\tt{phd2001141003@iiti.ac.in}}; \,\href{hkhungar@gmail.com}{\tt{hkhungar@gmail.com}}}}
\author[2]{Anirudh Singh Rana\thanks{\href{anirudh.rana@pilani.bits-pilani.ac.in}{\tt{anirudh.rana@pilani.bits-pilani.ac.in}}}}
\author[1]{Vinay Kumar Gupta\thanks{\href{vkg@iiti.ac.in}{\tt{vkg@iiti.ac.in}}}}
\affil[1]{Department of Mathematics, Indian Institute of Technology Indore,\newline Indore 453552, India}
\affil[2]{Department of Mathematics, Birla Institute of Technology and Science Pilani, Rajasthan 333031, India}
\date{}
\begin{document}
\begin{textblock*}{1.1\textwidth}(17.5mm,15mm)
\begin{center}
{\footnotesize
\color{green!50!black}
Published in the \href{https://doi.org/10.1103/PhysRevE.111.015101}{\ul{\emph{Phys.~Rev.~E}, {\bf 111}, 015101 (2025)}}
\\
A copy of the published version can also be obtained by mailing to the authors (emails are at the bottom of this page.)}
\end{center}
\end{textblock*}
\maketitle
\thispagestyle{empty}
\vspace*{-5mm}

\begin{abstract} 
{\color{black}
The well-known Navier--Stokes--Fourier equations of fluid dynamics are, in general, not adequate for describing rarefied gas flows.
Moreover, while the Stokes equations---a simplified version of the Navier--Stokes--Fourier equations---are effective in modeling slow and steady liquid flow
past a sphere,
they fail to yield a non-trivial solution to the problem of slow and steady liquid flow past an infinitely long cylinder (a two-dimensional problem essentially); this is referred to as Stokes' paradox.
The paradox also arises when studying these problems for gases. 
In this paper, we present a way to obtain meaningful solutions for two-dimensional flows of rarefied gases around objects by circumventing Stokes' paradox.
%
%
%
To this end, we adopt an extended hydrodynamic model, referred to as the CCR model, consisting of the balance equations for the mass, momentum and energy and closed with the coupled constitutive relations. 
We determine an analytic solution of the CCR model for the problem and compare it with a numerical solution based on the method of fundamental solutions. 
Apart from addressing flow past a circular cylinder, we aim to showcase the capabilities of the method of fundamental solutions to predict the flow past other objects in two dimensions for which analytic solutions do not exist or are difficult to determine. 
For that, we investigate the problem of rarefied gas flow past an infinitely long semicircular cylinder.

}
\end{abstract}

\section{\label{sec:intro}Introduction}
Fluid flow around stationary objects, especially spheres and cylinders, is a classic problem in fluid dynamics. 
Early research on low-speed viscous flows (often referred to as low-Reynolds-number flows) of incompressible fluids was pioneered by Sir George Gabriel Stokes in the 19th century. 
He postulated that at low velocities, the inertial forces become negligible with the pressure forces predominantly balanced by the viscous forces alone and, for such flows, the Navier--Stokes equations in turn boil down to the celebrated Stokes equations.
In honor of Sir Stokes, such a flow is referred to as a Stokes flow (or creeping flow).
Stokes flows are often encountered in nature, e.g., in swimming of microorganisms and sperms, and also in industries dealing with paints, polymers, etc. 

Stokes was successful in describing slow and steady flow of a viscous fluid past a sphere mathematically through the Stokes equations. 
However, when attempting to describe a slow and steady flow of a viscous fluid past an infinite cylinder (which is essentially a quasi-two-dimensional flow) using the Stokes equations~\citep{stokes1851}, he encountered difficulties in satisfying the boundary conditions at the cylinder surface and in the fluid at infinity simultaneously.
That led him to suggest the potential absence of a solution for the steady-state fluid flow past an infinite cylinder---a notion later coined as Stokes' paradox.

\begin{figure}[!t]
\centering
\includegraphics[scale=1]{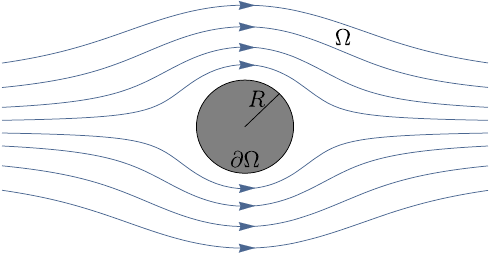} 
\caption{\label{stokes} Schematic of Stokes flow past an infinite circular cylinder of radius $R$, where the fluid is moving transversely to the axis of the cylinder.}
\end{figure}

To explain the paradox mathematically, we consider a viscous fluid moving slowly and steadily past an infinitely long right-circular cylinder of radius $R$ in the direction transverse to the axis of the cylinder as shown in Fig.~\ref{stokes}.
%
Let the flow domain be denoted by $\Omega$ and the boundary of the disk by $\partial\Omega$ and let the far-field velocity of the fluid be $(v_0,0,0)$ in the Cartesian coordinate system.
Owing to the symmetry around the axis of the cylinder, the problem essentially reduces to a (quasi-)two-dimensional problem or, equivalently, to the problem of fluid flow past a circular disk of radius $R$.
The Stokes equations for the problem read
%
%
\begin{align}
\label{StokesEqs}
\bm{\nabla}\cdot\bm{v}=0 \quad
\text{and} \quad \bm{\nabla} p -\mu \Delta\bm{v}=\bm{0} \quad\text{in}\quad \Omega,
\end{align}
where $\bm{v}$, $p$ and $\mu$ are the velocity, pressure and viscosity, respectively, of the fluid. 
The no-slip boundary condition on the surface of the cylinder reads
\begin{align}
\label{Stokes_noslip}
\bm{v}&=\bm{0} \quad\text{on}\quad \partial\Omega
\end{align}
and the far-field boundary condition reads 
\begin{align}
\label{Stokes_farfield}
\lim_{|x|\to \infty} v_x =v_0.
\end{align}
Eliminating the pressure $p$ from the Stokes equations \eqref{StokesEqs} and introducing the stream function $\psi(x,y)$---which is related to the components of the velocity via the relations $v_x=\partial\psi/\partial y$ and $v_y=-\partial\psi/\partial x$, the Stokes equations \eqref{StokesEqs} reduce to the biharmonic equation 
\begin{align}
\label{Stokes_polar}
\Delta^2\psi=0 \quad\text{in}\quad \Omega,  
\end{align}  
and the no-slip boundary condition \eqref{Stokes_noslip} reduces to 
\begin{align}
\label{Stokes_noslip_Cart}
\frac{\partial \psi}{\partial x} = 
\frac{\partial \psi}{\partial y} = 0
\quad\text{on}\quad \partial\Omega.
\end{align}
For determining $\psi$, it is convenient to transform the equations from the Cartesian coordinate system $(x,y)$ to the polar coordinate system $(r,\theta)$ so that $x=r \cos{\theta}$, $y=r \sin{\theta}$, $\psi(x,y) \equiv \psi(r,\theta)$, which is related to the components of the velocity in the polar coordinates via the relations $v_r=r^{-1}(\partial\psi/\partial\theta)$ and $v_\theta=-\partial\psi/\partial r$. 
In the polar coordinates, the no-slip boundary condition \eqref{Stokes_noslip_Cart} changes to 
\begin{align}
\label{Stokes_noslip_polar}
\frac{\partial \psi}{\partial r} = 
\frac{\partial \psi}{\partial \theta} = 0
\quad\text{at}\quad r=R \quad\text{and}\quad \forall \quad \theta\in[0,2\pi)
\end{align}
%
%
%
%
and the far-field condition \eqref{Stokes_farfield} changes to
\begin{align}
\label{Stokes_farfield_polar}
\lim_{r\to \infty}\frac{1}{r}\frac{\partial\psi}{\partial \theta}=v_0 \cos{\theta} \quad \text{and} \quad \lim_{r\to \infty}\frac{\partial\psi}{\partial r}=v_0 \sin{\theta}.
\end{align}
The far-field conditions \eqref{Stokes_farfield_polar} require that the stream function be of the form $\psi=f(r)\sin{\theta}$~\citep{Tanner1993, VanDyke1964, Lamb1932}.
Inserting this form of $\psi$ in (the polar form of) the biharmonic equation \eqref{Stokes_polar}, its solution reads~\citep{Tanner1993, Paloka2001}
\begin{align}
\label{strm_sol}
\psi(r,\theta)=\left(A r+\frac{B}{r}+C r^3+ D r \ln r\right) \sin{\theta},
\end{align}
where $A, B, C, D$ are the constants that need to be determined using boundary conditions \eqref{Stokes_noslip_polar} and \eqref{Stokes_farfield_polar}.
Both conditions in boundary condition~\eqref{Stokes_farfield_polar} imply that $C=D=0$ and $A=v_0$.
Consequently, there remains only one constant $B$  with which two conditions in boundary conditions~\eqref{Stokes_noslip_polar} are to be fulfilled, a scenario that is impossible unless $v_0=0$. 
This shows the nonexistence of solution to the Stokes equations for a steady flow past an infinite circular cylinder whereas such flows do exist physically---this is the essence of Stokes' paradox. 
The paradox arises not only in the case of flow past a circular cylinder, but also for an unbounded flow past any two-dimensional object of any shape \citep{Smith1990}. 

Another important consequence of Stokes' paradox is that the drag force on the cylinder in the aforementioned problem turns out to be infinite~\cite{Morra2018}, which is unreasonable physically.
Numerous endeavors have been dedicated to addressing Stokes' paradox and to determine the correct drag force on an infinitely long cylinder immersed in a viscous fluid moving transversely
to the axis of the cylinder.
Oseen~\citep{oseen1910}, in 1910, propounded an improvement to the Stokes equations by considering inertia effects at large distances and proposed the Oseen equations by adding convective acceleration terms to the Stokes equations. 
Oseen equations not only resolved Stokes' paradox but also led to an improved approximation of the drag force on a sphere immersed in a slow viscous flow.
Subsequent contributions by Lamb~\cite{lamb1911}, Bairstow~\emph{et al.}~\cite{BCL1923}, Tomotika and Aoi~\cite{TA1950}, refined the drag coefficient approximations for the cylinder using the Oseen equations. 
Further attempts to advance Oseen's ideas sparked the birth of a novel domain in applied mathematics known as the method of matched asymptotic expansions. 
Originally, Kaplun~\cite{Kaplun1954} and Kaplun and Lagerstrom~\cite{KL1957} executed the method of matched asymptotic expansions to obtain a new drag coefficient for flow past a circular cylinder. 
Further, Proudman and Pearson~\citep{PP1957} used the method of matched asymptotic expansions for flows past cylinder and sphere, and they came up with a novel drag result for sphere. 
Later, Kida and Take~\citep{KT1992}, through asymptotic expansions, provided expressions for the drag coefficient at different orders of approximation for low-Reynolds-number flow past a cylinder. 
Their results on the drag coefficient agreed well with experimental measurements at low Reynolds numbers. 
Recently, Khalili and Liu~\citep{KL2017} studied the problem of flow past a cylinder with the lattice-Boltzmann method and their simulation results on the drag coefficient led them to propose a slight correction to the expression for the drag coefficient obtained at the first order of approximation by~\citet{KT1992}.

%

In this paper, we revisit the problem of fluid flow past a cylinder but with fluid being a rarefied gas instead of a viscous liquid.
The reason for taking this problem is threefold: (i) the classic fluid dynamics models, e.g.~the Navier--Stokes--Fourier (NSF) equations (or Stokes equations for that matter), are incapable of capturing many intriguing rarefaction effects pertinent to rarefied gases, and hence better models (typically more involved than the Stokes equations, which are somewhat easy to handle) are needed for modeling rarefied gas flows, (ii) the occurrence of Stokes' paradox in rarefied gases as well poses mathematical challenges, and (iii) the problem leads to a method whose usefulness is noteworthy especially for problems for which an analytic solution either does not exist or is very difficult to obtain.
The problem is not only interesting from a theoretical point of view but research on rarefied gas flows past different objects is driven by the critical need to address challenges in space exploration, micro/nanotechnology, and vacuum system industries, etc.
Despite the presence of sufficient literature on two-dimensional unbounded flows in continuum fluid dynamics, there has been comparatively less attention towards rarefied gas flows past objects, particularly in two dimensions. 
To the best of the authors' knowledge, the first study on Stokes' paradox in rarefied gases was presented by \citet{Cercignani1968}, wherein he showed that, despite the fact that the Boltzmann equation is the most accurate model for investigating rarefied gases, the linearized Boltzmann equation---similarly to the Stokes equations---does not give bounded solutions for the flow past an axisymmetric body.
To circumvent Stokes' paradox, he proposed an inner-outer expansion of the Boltzmann equation.
\citet{YS1985} investigated rarefied gas flow past a circular cylinder at low Mach numbers by dividing the flow into two regions: (i) the kinetic region (flow domain near the cylinder)
modeling and (ii) near continuum region (flow domain outside the kinetic region). 
They handled the kinetic region with the simultaneous integral equations derived from the linearized Bhatnagar--Gross--Krook model and the  continuum region with the Oseen--Stokes equation. 
Their result on the drag on the cylinder matched reasonably well with those available in previous studies for a wide range of the Knudsen number. 
Recently, utilizing the advancements of moment methods in kinetic theory, \citet{GBJE2019} investigated non-equilibrium effects on flow past a circular cylinder. 
In the present paper, we utilize a relatively simple yet efficient model proposed by~\citet{RGS2018} that provides constitutive relations for the stress and heat flux appearing in the conservation laws. 
These relations are called the coupled constitutive relations (CCR) as the stress and heat flux in these relations are coupled with each other through a coupling coefficient and hence the model is referred to as the CCR model~\cite{RGS2018}. 
We aim to investigate and validate slow flow of a monatomic rarefied gas past an infinitely long right-circular cylinder using the CCR model.
To show the capabilities of our findings, we also  extend this study to rarefied gas flow past a semicircular cylinder.

To tackle the problem with the CCR model, we employ an efficient meshfree numerical technique, known as the method of fundamental solutions (MFS)~\citep{KA1964}.
The core idea of the MFS relies on the approximation of the solution of a (linear) boundary value problem  with the help of linear combination of free-space Green's functions or the fundamental solutions of the linear partial differential operator.
Once the fundamental solutions are known, the MFS is a very efficient numerical approach as it does not require domain discretization unlike the  oftentimes used numerical methods, such as finite volume and finite element methods that---at their core---rely on complex meshing to discretize the domain.
Owing to many advantages of the MFS over some other traditional numerical methods, the MFS has been applied to various problems in engineering and science~\citep{BK2001,FKM2003,KLM2011,LFS2021} and has recently gained popularity in exploring rarefied gas flows~\citep{LC2016,CSRSL2017,RSCLS2021,HRG2023}. 
Building on our previous work~\citep{HRG2023}, where we determined the fundamental solutions of the CCR model in two dimensions, in this paper we aim to extend  application of the MFS to rarefied gas flows past circular and semicircular cylinders (which are quasi-two-dimensional problems). 
This paper not only addresses Stokes' paradox but also validates the numerical scheme against analytic solution of the CCR model for the problem under consideration obtained using the symmetry-ansatz approach proposed by \citet{Torrilhon2010, WT2012}. 
To demonstrate the versatility and power of the MFS in solving complex rarefied gas flow problems, we extend the investigation to rarefied gas flow past a semicircular cylinder positioned at different angles for which no analytical solution exists or at least is very difficult to obtain.

The rest of the paper is structured as follows. The problem of rarefied gas flow past an infinitely long right-circular cylinder and the CCR model along with the boundary conditions are discussed in Sec.~\ref{sec:prob_form}. 
The procedure for finding an analytic solution of the CCR model for the problem under consideration is elucidated in Sec.~\ref{sec:analytical}, followed by the fundamental solutions of the CCR model and the technique to implement the MFS in Sec.~\ref{sec:implement}. 
The results for rarefied gas flow past an infinitely long circular cylinder are presented and validated in Sec.~\ref{sec:results1}. 
The problem of rarefied gas flow past an infinitely long semicircular cylinder is discussed in Sec.~\ref{sec:half}. 
The paper ends with the conclusion in Sec.~\ref{sec:conclusion}.

\section{\label{sec:prob_form}Problem formulation}
In this section, we describe the problem under consideration and the mathematical model with which the problem will be tackled.
Nonetheless, before this, let us point out that---to simplify the notations---all symbols with tilde as accent will henceforth denote quantities with dimensions while those without any accent will denote dimensionless quantities.

\subsection{\label{subsec:prob_disc}Problem description}
\begin{figure}[!b]
\centering
\includegraphics[scale=0.9]{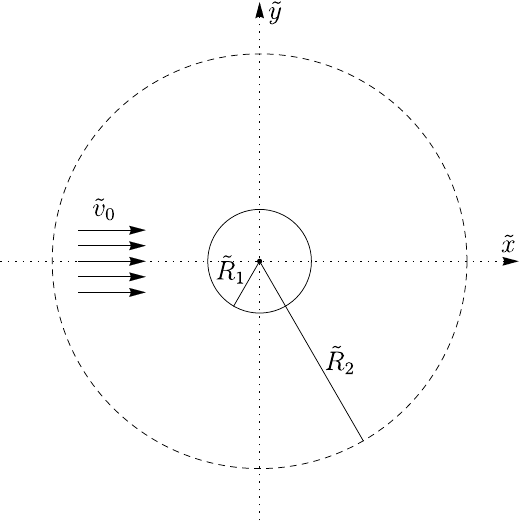} 
\caption{\label{fig:schematic1}Cross-sectional view of the problem of a rarefied gas flow past an infinitely long cylinder. 
The solid circle represents the periphery of the cylinder while the dashed circle represents an artificial boundary far away from the cylinder.}
\end{figure}

We consider a steady low-speed flow of a rarefied monatomic gas past an infinitely long right-circular cylinder having radius $\tilde{R}_1$. 
We assume that the cylinder is isothermal (having uniform temperature $\tilde{T}_0$, which is the same as the far-field ambient temperature of the gas) with large solid-to-gas thermal conductivity ratio. 
Let the circular cross section of the cylinder be in the $\tilde{x}\tilde{y}$-plane, the axis of the cylinder be coinciding with the $\tilde{z}$-axis and the flow be approaching the cylinder from the negative $\tilde{x}$-direction toward the positive $\tilde{x}$-direction.
As aforementioned, owing to the axial symmetry of the cylinder, the problem is quasi-two-dimensional, i.e., it is sufficient to study the problem for a circular disk of the same radius instead of studying the problem for the infinitely long cylinder.  
A two-dimensional cross-sectional view of the problem is depicted in Fig.~\ref{fig:schematic1} wherein the center of the disk is assumed to be fixed at the origin of the coordinate system. 
The radius of the disk is taken as the characteristic length scale $\tilde{L}$ for non-dimensionalization so that the dimensionless radius of the disk is $R_1=\tilde{R}_1/\tilde{L}=1$. 
To circumvent Stokes' paradox and hence the non-existence of a solution to the problem, we assume an artificial circular boundary of radius $\tilde{R}_2$ (where $\tilde{R}_2 \gg \tilde{R}_1$) outside the disk.
The radius of the artificial boundary is taken to be sufficiently large in comparison to the radius of the disk so that the artificial boundary has only insignificant effects on the problem under consideration.
%
\subsection{\label{sec:balance_eqs}Governing equations}
To account for rarefaction effects, we approach the problem with the CCR model~\citep{RGS2018} that consists of the mass, momentum and energy balance equations as the governing equations and special constitutive relations for the stress and heat flux which are coupled with each other.
Owing to the coupling between the constitutive relations for the stress and heat flux, the model has been referred to as the CCR model.
%
%
The governing equations (the mass, momentum and energy balance equations) in the CCR model reads~\citep{RGS2018}
\begin{gather}
\label{mass_bal}
\frac{\partial \tilde{\rho}}{\partial \tilde{t}}+\tilde{\bm{v}}\cdot\tilde{\bm{\nabla}}\tilde{\rho}+\tilde{\rho}\,\tilde{\bm{\nabla}}\cdot\tilde{\bm{v}}=0,
\\
\label{mom_bal}
\tilde{\rho}\left(\frac{\partial \tilde{\bm{v}}}{\partial \tilde{t}} + \tilde{\bm{v}}\cdot\tilde{\bm{\nabla}}\tilde{\bm{v}}\right) + \tilde{\bm{\nabla}} \tilde{p} + \tilde{\bm{\nabla}} \cdot \tilde{\bm{\sigma}} = \tilde{\rho} \tilde{\bm{F}},
\\
\label{energy_bal}
\tilde{\rho} \tilde{c}_v \left(\frac{\partial \tilde{T}}{\partial \tilde{t}} + \tilde{\bm{v}}\cdot\tilde{\bm{\nabla}} \tilde{T}\right) + \tilde{p} \, \tilde{\bm{\nabla}}\cdot\tilde{\bm{v}} + \tilde{\bm{\nabla}}\cdot\tilde{\bm{q}} + \tilde{\bm{\sigma}} :\tilde{\bm{\nabla}} \tilde{\bm{v}}=0,
\end{gather}
where $\tilde{\rho}$, $\tilde{\bm{v}}$, $\tilde{T}$, $\tilde{p}$, $\tilde{\bm{\sigma}}$, $\tilde{\bm{q}}$ are the density, velocity, temperature, pressure, stress tensor and heat flux, respectively; 
$\tilde{t}$ is the time variable;
$\tilde{\bm{F}}$ is the external force per unit mass; 
and the coefficient $\tilde{c}_v$ is the specific heat at constant volume and for monatomic gases $\tilde{c}_v = 3\tilde{R}/2$, with $\tilde{R}$ being the specific gas constant. 
Needless to say, the system of the governing equations \eqref{mass_bal}--\eqref{energy_bal} is not closed as such due to the presence of the additional unknowns $\tilde{\bm{\sigma}}$ and $\tilde{\bm{q}}$.
The constitutive relations in the CCR model for closing the system of equations \eqref{mass_bal}--\eqref{energy_bal} read~\citep{RGS2018}
\begin{align}
\label{ccr1}
\tilde{\bm{\sigma}}&=-2\tilde{\mu}\left[ \overline{\tilde{\bm{\nabla}}\tilde{\bm{v}}}
+  \frac{\alpha_0}{\tilde{p}} \bigg\{ \overline{\tilde{\bm{\nabla}}\tilde{\bm{q}}}
- \alpha_1 \, \overline{\tilde{\bm{q}}\,\tilde{\bm{\nabla}}(\ln \tilde{\theta})}
- \alpha_2 \, \overline{\tilde{\bm{q}}\,\tilde{\bm{\nabla}}(\ln \tilde{p})} \bigg\} \right],
\\
\label{ccr2}
\tilde{\bm{q}} &= - \tilde{\kappa}
\left[\tilde{\bm{\nabla}}\tilde{\theta}
+ \frac{\alpha_0}{\tilde{\rho}} \bigg\{ \tilde{\bm{\nabla}} \cdot \tilde{\bm{\sigma}}
-(1-\alpha_1) \, \tilde{\bm{\sigma}} \cdot\tilde{\bm{\nabla}} (\ln \tilde{\theta})
-(1-\alpha_2) \, \tilde{\bm{\sigma}} \cdot\tilde{\bm{\nabla}} (\ln \tilde{p})
\bigg\}\right],
\end{align}
where $\tilde{\mu}$ is the coefficient of the shear viscosity, $\tilde{\kappa}\tilde{R}$ is the coefficient of the thermal conductivity, $\tilde{\theta}=\tilde{R}\tilde{T}$ represents the temperature in energy units, and the overline above a quantity denotes its symmetric and tracefree part. 
For a $d$-dimensional vector $\bm{\psi}$, the symmetric-tracefree part of the tensor $\bm{\nabla} \bm{\psi}$ is defined as \cite{Gupta2020}
\begin{align}
\label{stf_tensor}
\overline{\bm{\nabla} \bm{\psi}}&=\frac{1}{2}\Big[{\bm{\nabla} \bm{\psi}}+(\bm{\nabla} \bm{\psi})^\mathsf{T}\Big]-\frac{1}{d}(\bm{\nabla}\cdot\bm{\psi})\bm{I},
\end{align}
where $\bm{I}$ is the identity tensor in $d$-dimensions.
For three- and quasi-two-dimensional problems, $d=3$.
The coefficient $\alpha_0$ in constitutive relations \eqref{ccr1} and \eqref{ccr2} is referred to as the coupling coefficient since it induces the coupling between constitutive relations for the stress and heat flux.
Setting $\alpha_0 = 0$ in \eqref{ccr1} and \eqref{ccr2} removes the coupling between the constitutive relations \eqref{ccr1} and \eqref{ccr2} and reduces them simply to the NSF constitutive relations.
The coefficients $\alpha_1$ and $\alpha_2$ in \eqref{ccr1} and \eqref{ccr2} are typically determined from experimental or theoretical scenarios; nonetheless, $\alpha_1=\alpha_2=0$ for Maxwell molecules~\citep{RGS2018}.
Equations~\eqref{mass_bal}--\eqref{energy_bal} along with the constitutive relations \eqref{ccr1} and \eqref{ccr2} are referred to as the CCR model~\citep{RGS2018}.

As we are interested in employing the MFS that relies on the linearity of equations, we shall be dealing with the linearized CCR model.
For linearization, we choose the equilibrium state of the gas as the reference state wherein let the density and temperature of the gas be $\tilde{\rho}_0$ and $\tilde{T}_0$, respectively, so that the pressure in the reference state be $\tilde{p}_0=\tilde{\rho}_0\tilde{\theta}_0$, where $\tilde{\theta}_0 = \tilde{R}\tilde{T}_0$. 
The other quantities (velocity, stress tensor and heat flux) in the reference state are zero.
For linearization, we introduce small perturbations in the flow variables from their values in the equilibrium state and, for convenience, we also make all quantities dimensionless using a length scale $\tilde{L}$, time scale $\tilde{L}/\sqrt{\tilde{\theta}_0}$ and appropriate combinations of the reference density $\tilde{\rho}_0$ and reference temperature $\tilde{T}_0$.
The dimensionless perturbations in the density, temperature, velocity, stress tensor and heat flux from their values in the reference state are given by
\begin{align}
\label{scalednoneqbquan}
\rho=\frac{\tilde{\rho}-\tilde{\rho}_0}{\tilde{\rho}_0},
\quad
T = \frac{\tilde{T}-\tilde{T}_0}{\tilde{T}_0},
\quad
\bm{v}=\frac{\tilde{\bm{v}}}{\sqrt{\tilde{\theta}_0}},
\quad 
\bm{\sigma}=\frac{\tilde{\bm{\sigma}}}{\tilde{\rho}_0 \tilde{\theta}_0}
\quad\text{and}\quad
\bm{q}=\frac{\tilde{\bm{q}}}{\tilde{\rho}_0(\tilde{\theta}_0)^{3/2}}, 
\end{align}
respectively.
Inserting these dimensionless perturbations in the CCR model \eqref{mass_bal}--\eqref{ccr2} and dropping all nonlinear terms in the dimensionless perturbations, we get the linear-dimensionless CCR model, which reads
\begin{gather}
\label{mass_bal_linear}
\frac{\partial \rho}{\partial t} + \bm{\nabla}\cdot \bm{v} = 0,
\\
\label{mom_bal_linear}
\frac{\partial \bm{v}}{\partial t}  + \bm{\nabla} p + \bm{\nabla} \cdot \bm{\sigma} = \bm{F},
\\
\label{energy_bal_linear}
c_v \frac{\partial T}{\partial t} 
+ \bm{\nabla} \cdot \bm{v} 
+ \bm{\nabla} \cdot \bm{q} = 0,
\end{gather}
\begin{align}
\label{ccr1_linear}
\bm{\sigma}&=-2\mathrm{Kn} 
\left( \overline{\bm{\nabla} \bm{v}}
+ \alpha_0 \, \overline{\bm{\nabla} \bm{q}}
 \right),
\\
\label{ccr2_linear}
\bm{q} &= - \frac{c_p \mathrm{Kn}}{\mathrm{Pr}}
\left(\bm{\nabla} T
+ \alpha_0 \, \bm{\nabla} \cdot \bm{\sigma}
\right),
\end{align}
where $t=\tilde{t}\sqrt{\tilde{\theta}_0}/\tilde{L}$, $\bm{\nabla} \equiv (1/\tilde{L})\tilde{\bm{\nabla}}$, the dimensionless perturbation in the pressure $p$ is $p = \rho + T$ due to the linearization, $c_v = \tilde{c}_v / \tilde{R}$,
\begin{align}
\label{KnPr}
\mathrm{Kn}=\frac{\tilde{\mu}_0}{\tilde{\rho}_0\sqrt{\tilde{\theta}_0}\tilde{L}}
\quad\text{and}\quad
\mathrm{Pr} = c_p \frac{\tilde{\mu}_0}{\tilde{\kappa}_0}
\end{align}
are the Knudsen number and Prandtl number, respectively, with $\tilde{\mu}_0$ and $\tilde{\kappa}_0 \tilde{R}$ being the coefficients of the shear viscosity and thermal conductivity, respectively, in the reference state.
In \eqref{KnPr}, $c_p = \tilde{c}_p / \tilde{R}$ with $\tilde{c}_p$ being the specific heat at constant pressure. 
For monatomic gases, $\tilde{c}_p = 5\tilde{R}/2$.
It may be noted that while performing the linearization, the external force $\tilde{\bm{F}}$ has been assumed to be small (of the order of perturbed variables) and has been scaled with $\tilde{\theta}_0/L$, i.e.~$\bm{F} = \tilde{\bm{F}} L/\tilde{\theta}_0$.
Equations~\eqref{mass_bal_linear}--\eqref{ccr2_linear} are referred to as the linear-dimensionless CCR model.

For the problem under consideration, the length scale $\tilde{L}$ is the radius of the disk $\tilde{R}_1$, there is no external force, i.e.~$\bm{F} = \bm{0}$, and the steady-state equations are obtained simply by setting all time-derivative terms in Eqs.~\eqref{mass_bal_linear}--\eqref{ccr2_linear} to zero, i.e. by setting $\partial(\cdot)/\partial{t} = 0$.
Consequently, the linear-dimensionless CCR model in the steady state reduces to
\begin{align}
\label{cons_laws}
\left.
\begin{aligned}
\bm{\nabla}\cdot\bm{v}=0,
\\
\bm{\nabla} p+\bm{\nabla}\cdot\bm{\sigma}=\bm{0},
\\
\bm{\nabla}\cdot\bm{q}=0,
\end{aligned}
\right\}
\end{align}
with the closure (for a monatomic gas)
\begin{align}
\label{ccr}
\left.
\begin{aligned}
\bm{\sigma}&=-2\mathrm{Kn} 
\left( \overline{\bm{\nabla} \bm{v}}
+ \alpha_0 \, \overline{\bm{\nabla} \bm{q}}
 \right),
\\
\bm{q} &= - \frac{c_p\mathrm{Kn}}{\mathrm{Pr}}
\left(\bm{\nabla} T
+ \alpha_0 \, \bm{\nabla} \cdot \bm{\sigma}
\right).
\end{aligned}
\right\}
\end{align}
It is worthwhile noting that for $\alpha_0=0$ the CCR model reduces to the NSF model and for $\alpha_0=2/5$ the steady-state linear-dimensionless CCR model (Eqs.~\eqref{cons_laws} and \eqref{ccr}) reduces to steady-state linear-dimensionless Grad $13$-moment equations; for more details, see \citep{RGS2018, RSCLS2021}. 
%
\subsection{\label{subsec:bc}Boundary conditions}
The thermodynamically-consistent boundary conditions complementing the linear CCR model were derived in Ref.~\cite{RSCLS2021} for evaporation/condensation problems and are given in Eqs.~($4.2$) and ($4.3$) of Ref.~\citep{RSCLS2021}. 
Since the problem under consideration does not involve evaporation or condensation, all of the Onsager reciprocity coefficients $\eta_{11}$, $\eta_{12}$ and $\eta_{22}$ in Eq.~($4.2$) of Ref.~\citep{RSCLS2021} are zero.
%
Moreover, the boundary conditions in Ref.~\citep{RSCLS2021} in a two-dimensional case are simplified further by considering only one tangential direction instead of two in Eq.~($4.3$) of Ref.~\citep{RSCLS2021}. 
Therefore, the boundary conditions complementing the linear CCR model in two dimensions are~\citep{RSCLS2021, HRG2023} 
\begin{align}
\label{bc_1}
(\bm{v}-\bm{v}_w)\cdot \bm{n}&=0,
\\
\label{bc_2}
\bm{q}\cdot \bm{n} &= -2\tau_0(T-T_w + \alpha_0 \, \bm{n}\cdot\bm{\sigma}\cdot\bm{n}),
\\
\label{bc_3}
\bm{t}\cdot\bm{\sigma}\cdot\bm{n} &= -\varsigma (\bm{v}-\bm{v}_w + \alpha_0 \, \bm{q})\cdot\bm{t},
\end{align}
where $\bm{n}$ and $\bm{t}$ are the unit normal and tangent vectors, respectively; and $\bm{v}_w$ and $\bm{T}_w$ are dimensionless perturbations in the velocity and temperature of the boundary wall. 
Note that the reference values of the velocity and temperature for making the velocity and temperature of the boundary wall linear and dimensionless are the same as those for making the velocity and temperature of the gas linear and dimensionless.
Equations \eqref{bc_2} and \eqref{bc_3} represent the temperature-jump and velocity-slip boundary conditions in which the temperature-jump and velocity-slip coefficients are given by \cite{RSCLS2021}
\begin{align}
\label{tj_vs}
\tau_0=0.8503\frac{\chi}{2-\chi}\sqrt{\frac{2}{\pi}}\quad\text{and}\quad 
\varsigma = 0.8798\frac{\chi}{2-\chi}\sqrt{\frac{2}{\pi}},
\end{align} 
respectively, where $\chi\in[0,1]$ is the accommodation coefficient that signifies the amount of the particles diffused (reflected) on (from) the wall into the gas. 
We assume the boundary to be diffusely reflecting, for which the accommodation coefficient $\chi=1$~\citep{Maxwell1879}. 
{\color{black}
The numerical factors in~\eqref{tj_vs} have been taken from Ref.~\cite{Sone2007} wherein they have been determined by performing an asymptotic expansion of the linearized Boltzmann equation in the limit $\mathrm{Kn}\to 0$.}

As pointed out in Sec.~\ref{sec:prob_form}, to circumvent Stokes' paradox, an artificial boundary in the flow domain has been assumed.  
To ensure that there is no disturbance to the flow due to this artificial boundary, the boundary conditions at the artificial boundary are taken as
\begin{align}
\label{bcs}
v_x=v_0,\quad
v_y=0\quad\text{and}\quad
T=0.
\end{align}
\section{\label{sec:analytical}Analytic solution}
As mentioned above, flow past an infinitely long right circular cylinder is actually a quasi-two-dimensional problem when the fluid flow is in the normal direction of the axis of the cylinder.
In this case, there is no change in the flow variables in the axial direction of the cylinder. 
To tackle the problem, it is convenient to work in a cylindrical coordinate system $(r,\vartheta,z)$, wherein the $z$-axis coincides with the axis of the cylinder.
Owing to the axial symmetry, the flow variables do not change in the $z$-direction.

In this cylindrical coordinate system, the linear steady-state CCR model (Eqs.~\eqref{cons_laws} and \eqref{ccr}) can be written as follows.
%
The mass, momentum and energy balance equations \eqref{cons_laws} in the cylindrical coordinate system read
\begin{align}
\label{massbal_cyl}
\frac{\partial v_r}{\partial r}+\frac{1}{r}\frac{\partial v_\vartheta}{\partial \vartheta}+\frac{v_r}{r}&=0,
\\
\newsubeqblock
\mysubeq
\label{r_mombal_cyl}
\frac{\partial p}{\partial r}+\frac{\partial \sigma_{rr}}{\partial r}+\frac{1}{r}\frac{\partial \sigma_{r \vartheta}}{\partial\vartheta}+\frac{\sigma_{rr}-\sigma_{\vartheta\vartheta}}{r}&=0,
\\
\mysubeq
\label{theta_mombal_cyl}
\frac{1}{r}\frac{\partial p}{\partial \vartheta}+\frac{\partial \sigma_{r\vartheta}}{\partial r}+\frac{1}{r}\frac{\partial \sigma_{\vartheta\vartheta}}{\partial\vartheta}+\frac{2\sigma_{r\vartheta}}{r}&=0,
\\
\label{energybal_cyl}
\frac{\partial q_r}{\partial r}+\frac{1}{r}\frac{\partial q_\vartheta}{\partial \vartheta}+\frac{q_r}{r}&=0,
\end{align}
where \eqref{massbal_cyl} is the mass balance equation \eqref{cons_laws}$_1$, \eqref{r_mombal_cyl} and \eqref{theta_mombal_cyl} are the momentum balance equation \eqref{cons_laws}$_2$ in the $r$- and $\theta$-directions, respectively, and \eqref{energybal_cyl} is the energy balance equation \eqref{cons_laws}$_3$. 
It may be noted that the momentum balance equation in the $z$-direction is trivially satisfied, owing to the fact that there is no change in flow variables with respect to the $z$-coordinate.
The required variables to close the system \eqref{massbal_cyl}--\eqref{energybal_cyl} from the closure relations \eqref{ccr}---in the cylindrical coordinate system---read
%
\begin{align}
\newsubeqblock
\mysubeq
\sigma_{rr}&=-2\mathrm{Kn}\frac{\partial v_r}{\partial r}-2\mathrm{Kn}\alpha_0\frac{\partial q_r}{\partial r},
\\
\mysubeq
\sigma_{r\vartheta}&=-\mathrm{Kn}\left(\frac{\partial v_\vartheta}{\partial r}+\frac{1}{r}\frac{\partial v_r}{\partial \vartheta}-\frac{v_\vartheta}{r}\right)-\alpha_0\mathrm{Kn}\left(\frac{\partial q_\vartheta}{\partial r}+\frac{1}{r}\frac{\partial q_r}{\partial \vartheta}-\frac{q_\vartheta}{r}\right),
\\
\mysubeq
\sigma_{\vartheta\vartheta}&=-2\mathrm{Kn}\left(\frac{1}{r}\frac{\partial v_\vartheta}{\partial \vartheta}+\frac{v_r}{r}\right)-2\alpha_0\mathrm{Kn}\left(\frac{1}{r}\frac{\partial q_\vartheta}{\partial \vartheta}+\frac{q_r}{r}\right),
\\
\newsubeqblock 
\mysubeq
\label{qradial}
q_r&=-\frac{c_p \mathrm{Kn}}{\mathrm{Pr}}\left[\frac{\partial T}{\partial r}+\alpha_0\left(\frac{\partial \sigma_{rr}}{\partial r}+\frac{1}{r}\frac{\partial \sigma_{r \vartheta}}{\partial\vartheta}+\frac{\sigma_{rr}-\sigma_{\vartheta\vartheta}}{r}\right)\right],
\\
\label{cyl2}
\mysubeq
q_\vartheta&=-\frac{c_p \mathrm{Kn}}{\mathrm{Pr}}\left[\frac{1}{r}\frac{\partial T}{\partial \vartheta}+\alpha_0\left(\frac{\partial \sigma_{r\vartheta}}{\partial r}+\frac{1}{r}\frac{\partial \sigma_{\vartheta\vartheta}}{\partial\vartheta}+\frac{2\sigma_{r\vartheta}}{r}\right)\right].
\end{align}
To determine an analytic solution of the CCR model \eqref{massbal_cyl}--\eqref{cyl2} (in quasi-two dimensions), we convert the partial differential equations \eqref{massbal_cyl}--\eqref{cyl2} into ordinary differential equations using symmetry ansatz, which is inspired by the solution of the Stokes equations.
%
This approach has also been utilized to determine analytic solutions of the regularized 13-moment (R13) and regularized 26-moment (R26) equations in the linearized state for the problems of flow past a sphere and a cylinder~\citep{Torrilhon2010, WT2012, RGST2021}. 
In symmetry ansatz, the radial dependency of the variables is separated and the angular dependency of the variables is expressed using the sine and cosine functions. 
For this purpose, the vector and tensor components having an odd number of indices in $\vartheta$ are selected to be proportional to $\sin \vartheta$ whereas the scalars and tensor components with an even number of indices in $\vartheta$ are made proportional to $\cos\vartheta$~\citep{Torrilhon2010}. 
Furthermore, since the problem is quasi-two-dimensional, the $z$-coordinate dependency of the variables is automatically eliminated.
With these symmetry ansatz, the solution for the vectors $\bm{v}$ and $\bm{q}$ should be of the form
\begin{align}
\label{vec_ansatz}
\bm{v}(r,\vartheta)=\begin{bmatrix}
a(r)\,\cos{\vartheta}\\
b(r)\,\sin{\vartheta}\\ 0
\end{bmatrix}
\quad\text{and}\quad
\bm{q}(r,\vartheta)=\begin{bmatrix}
\alpha(r)\,\cos{\vartheta}\\
\beta(r)\,\sin{\vartheta}\\ 0
\end{bmatrix},
\end{align}
that for the scalars $p$ and $T$ should be of the form
\begin{align}
\label{sca_ansatz}
p(r,\vartheta)=c(r)\cos{\vartheta}
\quad\text{and}\quad
T(r,\vartheta)=d(r)\cos{\vartheta},
\end{align}
and that for $\bm{\sigma}$ should be of the form 
\begin{align}
\label{ten_ansatz}
\bm{\sigma}(r,\vartheta)=\begin{bmatrix}
\gamma(r)\,\cos{\vartheta} & \kappa(r)\,\sin{\vartheta} & 0\\
\kappa(r)\,\sin{\vartheta} & \omega(r)\,\cos{\vartheta} & 0\\
0 & 0 & \sigma_{zz}
\end{bmatrix},
\end{align}
where $a(r)$, $b(r)$, $\alpha(r)$, $\beta(r)$, $c(r)$, $d(r)$, $\gamma(r)$, $\kappa(r)$ and $\omega(r)$ are the unknown functions that need to be determined, and $\sigma_{zz} = - \big[\gamma(r) + \omega(r)\big]\cos{\vartheta}$ as $\bm{\sigma}$ is a symmetric and tracefree tensor of rank 2.
Insertion of ansatz \eqref{vec_ansatz}--\eqref{ten_ansatz} in Eqs.~\eqref{massbal_cyl}--\eqref{cyl2} leads to a system of ordinary differential equations in the unknowns $a(r)$, $b(r)$, $\alpha(r)$, $\beta(r)$, $c(r)$, $d(r)$, $\gamma(r)$, $\kappa(r)$ and $\omega(r)$ that is solved to determine these unknowns.
%
For the sake of brevity, we omit presenting the explicit values of these unknowns here. 
Nevertheless, substituting the obtained values of the unknowns in ansatz \eqref{vec_ansatz}--\eqref{ten_ansatz}, we get the following solution for the field variables.
\begingroup
\allowdisplaybreaks
\begin{align}
\label{vr}
v_r(r,\vartheta)&=\left(c_3-\frac{c_4}{r^2}+c_5 r^2+c_6\ln{r}\right)\cos{\vartheta},
\\
\label{vt}
v_\vartheta(r,\vartheta)&=\left(-c_3-\frac{c_4}{r^2}-3c_5 r^2-c_6-c_6\ln{r}\right)\sin{\vartheta},
\\
\label{qr}
q_r(r,\vartheta)&=\left(\frac{c_1}{r^2}+c_2\right)\cos{\vartheta},
\\
\label{qt}
q_\vartheta(r,\vartheta)&=\left(\frac{c_1}{r^2}-c_2\right)\sin{\vartheta},
\\
\label{p}
p(r,\vartheta)&=\mathrm{Kn}\left(8r c_5-\frac{2c_6}{r}\right)\cos{\vartheta},
\\
\label{srr}
\sigma_{rr}(r,\vartheta)&=\mathrm{Kn}\left(\frac{4\alpha_0 c_1}{r^3}-\frac{4c_4}{r^3}-4r c_5-\frac{2c_6}{r}\right)\cos{\vartheta},
\\
\label{srt}
\sigma_{r\vartheta}(r,\vartheta)&=\mathrm{Kn}\left(\frac{4\alpha_0 c_1}{r^3}-\frac{4c_4}{r^3}+4r c_5\right)\sin{\vartheta},
\\
\label{stt}
\sigma_{\vartheta\vartheta}(r,\vartheta)&=\mathrm{Kn}\left(\frac{-4\alpha_0 c_1}{r^3}+\frac{4c_4}{r^3}+4r c_5+\frac{2c_6}{r}\right)\cos{\vartheta},
\\
\label{t}
T(r,\vartheta)&=\left(\frac{
\mathrm{Pr}}{c_p \mathrm{Kn}}\left(\frac{c_1}{r}-rc_2\right)+\mathrm{Kn}\alpha_0\left(8rc_5-\frac{2c_6}{r}\right)\right)\cos{\vartheta}.
\end{align}
\endgroup
The constants $c_1$, $c_2$, $c_3$, $c_4$, $c_5$ and $c_6$ in the above solution are determined using boundary conditions \eqref{bc_1}--\eqref{bc_3} and \eqref{bcs}. 
However, it is crucial to acknowledge that without the presence of the outer artificial wall or, in other words, without imposing the specified boundary conditions \eqref{bcs}, determining the six constants $c_1$, $c_2$, $c_3$, $c_4$, $c_5$ and $c_6$ uniquely through three boundary conditions \eqref{bc_1}--\eqref{bc_3} is impracticable. 
Moreover, for solutions to converge in the far field (as $r\to \infty$), it becomes necessary that the constants $c_2=c_5=c_6=0$. 
Additionally, if the boundary conditions \eqref{bc_1}--\eqref{bc_3} are imposed, it follows that the remaining constants $c_1, c_3,$ and $c_4$ also become zero, resulting in an overall zero solution. 
This scenario illustrates the occurrence of Stokes' paradox with the CCR model as well, and thereby affirms the necessity of employing an artificial boundary to circumvent this paradox.
Therefore, the constants $c_1,c_2,\dots,c_6$ are determined using boundary conditions \eqref{bc_1}--\eqref{bc_3} and \eqref{bcs}. 
Substituting the determined values of the constants, the final flow variables---when required---can be converted back into the Cartesian coordinate system using the transformation
\begin{align}
\begin{bmatrix}
\hat{x}\\
\hat{y}\\
\hat{z}
\end{bmatrix}=
\begin{bmatrix}
\cos{\vartheta} & -\sin{\vartheta}  & 0\\
\sin{\vartheta} & \hphantom{-}\cos{\vartheta} & 0\\
0  & \hphantom{-}0 & 1
\end{bmatrix}\begin{bmatrix}
\hat{r}\\
\hat{\vartheta}\\ \hat{z}
\end{bmatrix},
\end{align}
where $\hat{x},\hat{y},\hat{z}$ denote the unit vectors in the Cartesian coordinate system and $\hat{r},\hat{\vartheta},\hat{z}$ are the unit vectors in the polar coordinate system. 
For instance, the velocity is given by
\begin{align}
\bm{v}=(v_x,v_y,0)^\mathsf{T}=(v_r \cos{\vartheta} - v_\vartheta \sin{\vartheta}, v_r \sin{\vartheta} + v_\vartheta\cos{\vartheta},0 )^\mathsf{T}.
\end{align}
%
\section{\label{sec:implement}Fundamental solutions and their implementation}
The MFS utilizes the fundamental solutions (or free-space Green's functions) of the governing equations as the basis functions. 
The fundamental solutions have their own source points (or singularities) located outside the physical domain of the problem, ensuring that the MFS remains free from singularities.
In approximating the numerical solution through the superposition of the fundamental solutions, the unknowns associated with the source points are introduced and these unknowns are determined by satisfying the boundary conditions at the discretized boundary nodes (or collocation points). 
Consequently, the overall solution relies on the unknowns related to the source points and on the boundary conditions for the problem.
The two-dimensional fundamental solutions for the CCR model have been derived by us in Ref.~\citep{HRG2023}, wherein it has been shown that
the fundamental solutions of the CCR model depend on the point mass source $h$, point force $\bm{f}=(f_1,f_2)^\mathsf{T}$ and point heat source $g$ that were included as the Dirac delta forcing terms in the mass, momentum and energy balance equations, respectively. 
The point mass source $h$ considered in Eq.~(20) of Ref.~\citep{HRG2023} is nonzero only for the problems involving evaporation or condensation. 
In the present paper, we do not deal with such problems, so it is not necessary to include the forcing term in the mass balance equation, i.e.~the governing equations with the forcing terms for the problem under consideration read
\begin{align}
\label{cons_laws_with_sources}
\left.
\begin{aligned}
\bm{\nabla}\cdot\bm{v}&=0,
\\
\bm{\nabla} p+\bm{\nabla}\cdot\bm{\sigma}&=\bm{f}\,\delta(\bm{r}),
\\
\bm{\nabla}\cdot\bm{q}&=g\,\delta(\bm{r}),
\end{aligned}
\right\}
\end{align}
where $\bm{r}$ is the position of the singularity whereat the point sources $\bm{f}$ and $g$ are placed.
Solving system \eqref{cons_laws_with_sources} with constitutive relations~\eqref{ccr} yields the fundamental solutions of the CCR model in two dimensions. 
A method of determining the fundamental solutions of the CCR model in two dimensions is given in Ref.~\citep{HRG2023}.
But for the sake of completeness, we also present a slightly different way of determining these fundamental solutions in Appendix~\ref{app:A}.
Notwithstanding, the methods given in Appendix~\ref{app:A} and in Ref.~\cite{HRG2023} are equivalent and give exactly the same fundamental solutions.
The fundamental solutions of the CCR model in two dimensions read~\citep{HRG2023} (see also Appendix~\ref{app:A})
\begin{align}
\label{vel}
\bm{v}(\bm{r})&=\frac{1}{8\pi \mathrm{Kn}}\bm{f}\cdot\left[\frac{2\bm{rr}}{r^2}-(2\ln{r}-1)\bm{I}\right]+\frac{c_p \mathrm{Kn}}{2\pi \mathrm{Pr}}\alpha_0^2 \bm{f}\cdot \left(\frac{2\bm{rr}}{r^4}-\frac{\bm{I}}{r^2}\right),
\\
\label{pressure}
p(\bm{r})&=\frac{\bm{f\cdot r}}{2\pi r^2},
\\
\label{stress}
\bm{\sigma}(\bm{r})&=\frac{2 \mathrm{Kn}\, g\,\alpha_0 +\bm{f\cdot r}}{2\pi} \left(\frac{2\bm{rr}}{r^4}-\frac{\bm{I}}{r^2}\right),
\\
\label{temp}
T(\bm{r})&=-\frac{\mathrm{Pr} \,g}{2 \pi \mathrm{Kn} \,c_p} \ln{r},
\\
\label{heat}
\bm{q}(\bm{r})&=\frac{g}{2\pi}\frac{\bm{r}}{r^2}-\frac{c_p \mathrm{Kn}}{2\pi\, \mathrm{Pr}}\alpha_0 \bm{f}\cdot \left(\frac{2\bm{rr}}{r^4}-\frac{\bm{I}}{r^2}\right),
\end{align}
where $r=|\bm{r}|$.

\begin{figure}[!tb]
\centering
\includegraphics[scale=0.9]{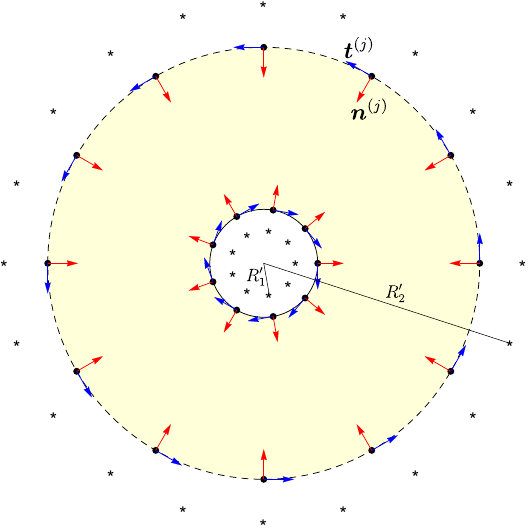}
\caption{\label{schematic_sing}Placement of the collocation points (black dots) on the boundary and singularities (black stars) outside the confined domain. The blue and red arrows at each boundary node denote the unit tangent and normal vectors, respectively.} 
\end{figure}

As discussed in Sec.~\ref{sec:intro}, the mathematical origin of Stokes' paradox lies in the logarithmic dependence of the solution of the Stokes equations. 
This logarithmic dependence is also seen in the fundamental solutions of the CCR model (see Eqs.~\eqref{vel} and \eqref{temp}), due to which the solution diverges in the far field. 
To circumvent this difficulty, we employ the MFS on a bounded domain by again introducing an artificial outer boundary which is far enough from the original circular disk. 
To place the singularity points outside the computational domain, we assume that the source points are located on two circles---one inside the actual periphery of the disk and the other outside of the artificial boundary. 
The circles on which the singularity points are placed will henceforth be referred to as the fictitious boundaries.
An illustration depicting the boundary nodes on the periphery of the disk and on the artificial boundary, and the location of source points on the fictitious boundaries is presented in Fig.~\ref{schematic_sing}.
For more details on the location of source points, the reader is referred to Ref.~\cite{HRG2023}. 
We consider a total of $N_s$ source points, out of which $N_{s_1}$ points lie on the inner fictitious boundary having dimensionless radius $R_1^\prime$ and $N_{s_2}$ points on the outer fictitious boundary having dimensionless radius $R_2^\prime$ (where $R_1^\prime < R_1$ and $R_2^\prime > R_2$).
Additionally, we place $N_{b_1}$ boundary nodes on the actual periphery of the disk and $N_{b_2}$ boundary nodes on the artificial boundary accounting for a total of $N_b$ boundary points. 
To define the positions, we denote the position of the $i^\mathrm{th}$ singularity by $\bm{x}_s^{(i)}$ and that of the $j^\mathrm{th}$ boundary node by $\bm{x}_b^{(j)}$.  
The vector from the $i^\mathrm{th}$ singularity site to the point situated at position $\bm{x}$ in the domain is represented by $\bm{r}^{(i)}=\bm{x}-\bm{x}_s^{(i)}$
and the vector from the $i^\mathrm{th}$ singularity site to the $j^\mathrm{th}$ boundary node is denoted by $\bm{r}^{(ij)}=\bm{x}_b^{(j)}-\bm{x}_s^{(i)}$.
Corresponding to the $i^\mathrm{th}$ singularity ($i=1,2,3,\dots,N_s$), there are three unknowns, namely $f_1^{(i)}$, $f_2^{(i)}$ and $g^{(i)}$, where $f_1^{(i)}$ and $f_2^{(i)}$ are the components of the point force $\bm{f}^{(i)}$ applied on the $i^\mathrm{th}$ singularity, i.e.~$\bm{f}^{(i)} = \Big(f_1^{(i)}, f_2^{(i)}\Big)^\mathsf{T}$, and $g^{(i)}$ is the point heat source applied on the $i^\mathrm{th}$ singularity. 
Thus there are a total of $3\times N_s$ unknowns. 
These unknowns are to be computed using the boundary conditions at both the actual and artificial boundaries. 
This means that three boundary conditions need to be applied at each boundary node, which leads to a set of $3\times N_b$ linear algebraic equations that are to be solved for $3\times N_s$ unknowns. 
Since the boundary conditions are to be applied on the boundary nodes, it makes sense to write the field variables---obtained from the superposition of the fundamental solutions \eqref{vel}--\eqref{heat} for all singularities---at the $j^\mathrm{th}$ boundary node.
On superposition of the fundamental solutions \eqref{vel}--\eqref{heat} for all singularities, the field variables at the $j^\mathrm{th}$ boundary node read
\begin{align}
\label{vel_j}
\bm{v}^{(j)}&=\sum_{i=1}^{N_s}\frac{\bm{f}^{(i)}}{8\pi \mathrm{Kn}}\cdot\left(\frac{2\bm{r}^{(ij)}\bm{r}^{(ij)}}{|\bm{r}^{(ij)}|^2}-(2\ln{|\bm{r}^{(ij)}|}-1)\bm{I}\right)+\frac{c_p \mathrm{Kn}\,\alpha_0^2}{2\pi \mathrm{Pr}} \bm{f}^{(i)}\cdot \left(\frac{2\bm{r}^{(ij)}\bm{r}^{(ij)}}{|\bm{r}^{(ij)}|^4}-\frac{\bm{I}}{|\bm{r}^{(ij)}|^2}\right),
\\
p^{(j)}&=\sum_{i=1}^{N_s}\frac{\bm{f}^{(i)}\cdot \bm{r}^{(ij)}}{2\pi |\bm{r}^{(ij)}|^2},
\\
\bm{\sigma}^{(j)}&=\sum_{i=1}^{N_s}\frac{2 \mathrm{Kn}\,g^{(i)}\,\alpha_0+\bm{f}^{(i)}\cdot \bm{r}^{(ij)}}{2\pi} \bm{K}(\bm{r}^{(ij)}),
\\
T^{(j)}&=-\sum_{i=1}^{N_s}\frac{ g^{(i)}\, \mathrm{Pr}}{c_p \mathrm{Kn}}\frac{\ln{|\bm{r}^{(ij)}|}}{2\pi},
\\
\label{heat_j}
\bm{q}^{(j)}&=\sum_{i=1}^{N_s}\left(\frac{g^{(i)}}{2\pi}\frac{\bm{r}^{(ij)}}{|\bm{r}^{(ij)}|^2}-\frac{c_p \mathrm{Kn}}{2\pi \mathrm{Pr}}\alpha_0 \bm{f}^{(i)}\cdot \left(\frac{2\bm{r}^{(ij)}\bm{r}^{(ij)}}{|\bm{r}^{(ij)}|^4}-\frac{\bm{I}}{|\bm{r}^{(ij)}|^2}\right) \right).
\end{align}
From boundary conditions \eqref{bc_1}--\eqref{bc_3}, the boundary conditions for the $j^{\mathrm{th}}$ boundary node on the actual periphery of the disk are
\begin{align}
\label{vn}
\bm{v}^{(j)}\cdot\bm{n}^{(j)}&=0,
\\
\label{temp_jump}
\bm{q}^{(j)}\cdot\bm{n}^{(j)}+2\tau_0\left[T^{(j)}+\alpha_0\bm{n}^{(j)}\cdot\bm{\sigma}^{(j)}\cdot\bm{n}^{(j)}\right]&=0,
\\
\label{vel_slip}
\bm{n}^{(j)}\cdot\bm{\sigma}^{(j)}\cdot\bm{t}^{(j)}+\varsigma\left[\bm{v}^{(j)}+\alpha_0\bm{q}^{(j)}\right]\cdot\bm{t}^{(j)}&=0
\end{align}
for $j=1,2,\dots ,N_{b_1}$. 
Here, $\bm{n}^{(j)}$ and $\bm{t}^{(j)}$ are the unit vectors normal and tangent to the boundary at the $j^{\mathrm{th}}$ boundary node.
From boundary conditions \eqref{bcs}, the boundary conditions for the $j^{\mathrm{th}}$ boundary node on the artificial boundary are 
\begin{align}
\label{outer_bc}
v_x^{(j)}=v_0,\quad
v_y^{(j)}=0\quad\text{and}\quad
T^{(j)}=0
\end{align}
for $j=1,2,\dots,N_{b_2}$. 

Now, using the field variables at the $j^{\mathrm{th}}$ boundary node from Eqs.~\eqref{vel_j}--\eqref{heat_j} in boundary conditions \eqref{vn}--\eqref{vel_slip} for $j=1,2,\dots ,N_{b_1}$ and in boundary conditions \eqref{outer_bc} for $j=1,2,\dots ,N_{b_2}$, we obtain a system of $3N_{b_1}+3N_{b_2}=3 N_b$ linear equations in $3N_s$ unknowns, namely $f_1^{(1)}$, $f_2^{(1)}$, $g^{(1)}$, $f_1^{(2)}$, $f_2^{(2)}$, $g^{(2)}$,\dots, $ f_1^{(N_s)}$, $f_2^{(N_s)}$, $g^{(N_s)}$. 
This system can be written in a matrix form as
\begin{align}
\label{mat_eq}
\mathcal{M} \bm{U} = \bm{b},
\end{align}
where $\bm{U}$ is the column vector containing all the unknowns, i.e.~$\bm{U} = \Big(f_1^{(1)},f_2^{(1)},g^{(1)}
, f_1^{(2)},f_2^{(2)},g^{(2)},\allowbreak\dots,
 f_1^{(N_s)},f_2^{(N_s)},g^{(N_s)}\Big)^\mathsf{T}$; $\mathcal{M}$ is the corresponding coefficient matrix, often referred to as the collocation matrix; and $\bm{b}$ is the column vector containing only the wall properties, e.g.~$v_0$.
We have solved the system using the method of least squares in Mathematica. 
Since the MFS may lead to a bad-conditioned collocation matrix, it is favorable to use the method of least squares even if the collocation matrix is square~\cite{CH2020}.
The obtained unknowns facilitate the computation of the flow properties across the entire domain as the flow variables at position $\bm{x}$ in the flow domain can be determined simply by
dropping the superscript ``$(j)$" and ``$j$" from the superscript ``$(ij)$" everywhere in Eqs.~\eqref{vel_j}--\eqref{heat_j}. 
For example, the velocity $\bm{v} \equiv \bm{v}(\bm{x})$ at position $\bm{x}$ in the flow
domain is given by
\begin{align}
\bm{v} = \sum_{i=1}^{N_s}\frac{\bm{f}^{(i)}}{8\pi \mathrm{Kn}}\cdot\left(\frac{2\bm{r}^{(i)}\bm{r}^{(i)}}{|\bm{r}^{(i)}|^2}-(2 \ln{|\bm{r}^{(i)}|}-1)\bm{I}\right)+\frac{c_p \mathrm{Kn}\,\alpha_0^2}{2\pi \mathrm{Pr}} \bm{f}^{(i)}\cdot \left(\frac{2\bm{r}^{(i)}\bm{r}^{(i)}}{|\bm{r}^{(i)}|^4}-\frac{\bm{I}}{|\bm{r}^{(i)}|^2}\right).
\end{align}
%


\section{\label{sec:results1}Results and discussion for flow past a circular cylinder}
For numerical computations, we fix the dimensionless radius of the artificial boundary to $R_2=10$ and the dimensionless radii of the inner and outer fictitious boundaries to $R_1^\prime=0.5$ and $R_2^\prime=20$, respectively, 
the number of boundary nodes on the actual periphery of the disk to $N_{b_1}=50$ and the number of boundary nodes on the artificial boundary to $N_{b_2}=100$.
For simplicity, we fix the number of singularity points on the inner fictitious boundary to be equal to the number of inner boundary nodes, i.e.~$N_{s_1}=N_{b_1}=50$, and the number of singularity points on the outer fictitious boundary to be equal to the number of boundary nodes on the outer artificial boundary, i.e.~$N_{s_2}=N_{b_2}=100$ so as to make the collocation matrix (having dimensions $3 N_b\times 3 N_s$) square.
Notwithstanding, the results obtained with a rectangular collocation matrix do not differ significantly from those obtained with a square collocation matrix in the present paper since we have used the method of least squares for solving the formed system of equations numerically. 
Furthermore, the (dimensionless) approaching velocity (in the $x$-direction) of the gas far away from the cylinder has been fixed to $v_0=1$.
%

\begin{figure}[!tb]
\centering
\includegraphics[scale=0.49]{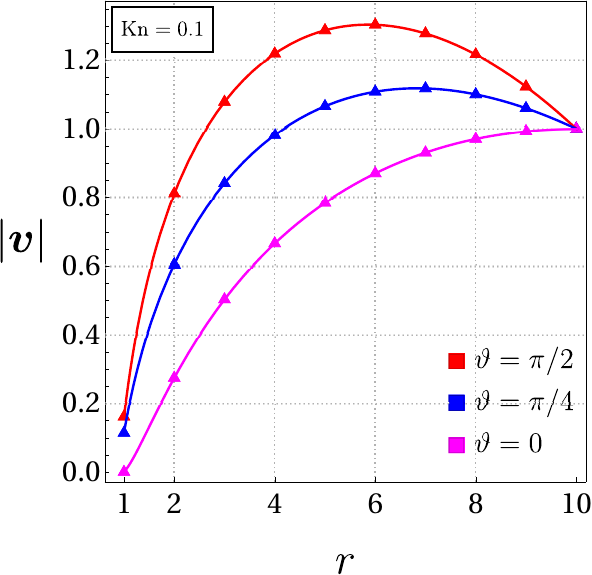}\quad \includegraphics[scale=0.49]{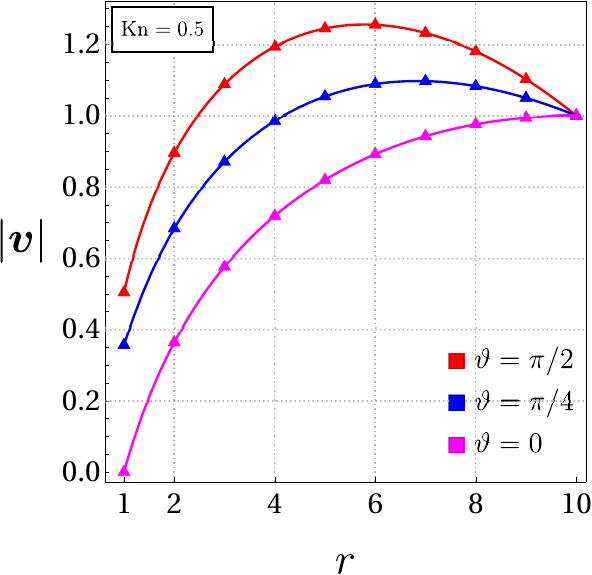} \quad
\includegraphics[scale=0.49]{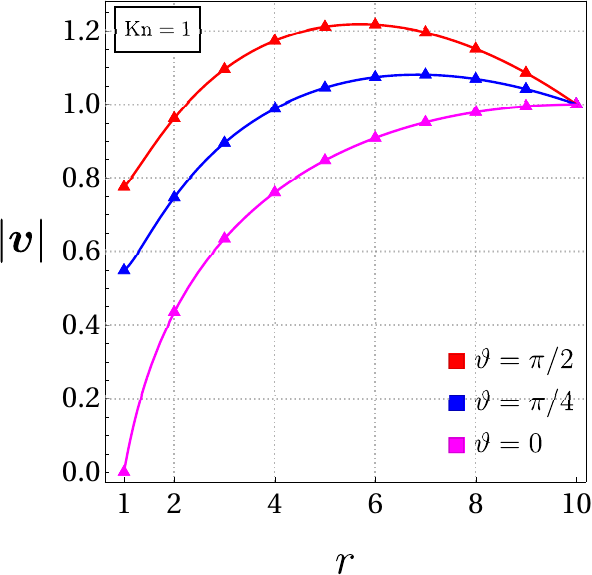}
\caption{\label{fig:speed}Speed of the gas varying with the radial position in different directions for $\mathrm{Kn}=0.1$, $0.5$ and $1$. The solid lines represent the results obtained from the MFS applied to the CCR model and the symbols represent the analytic solutions. 
The other parameters are $N_{b_1}=N_{s_1}=50$, $N_{b_2}=N_{s_2}=100$, $R_1=1$, $R_2=10$, $R_1^\prime=0.5$, $R_2^\prime=20$.}
\end{figure}

\begin{figure}[!tb]
\centering
\includegraphics[scale=0.35]{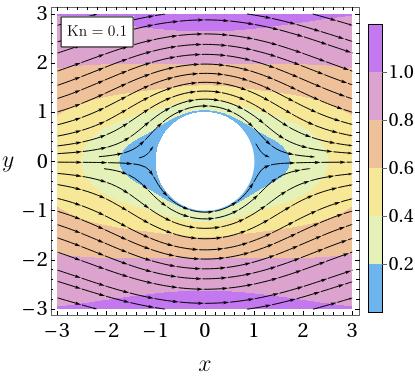} \includegraphics[scale=0.35]{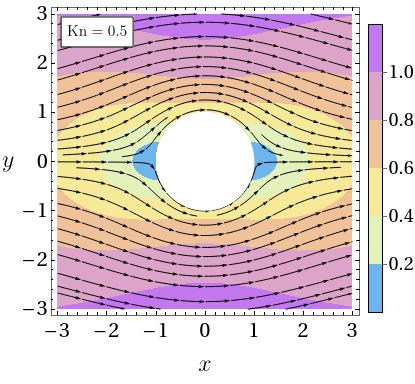} 
\includegraphics[scale=0.35]{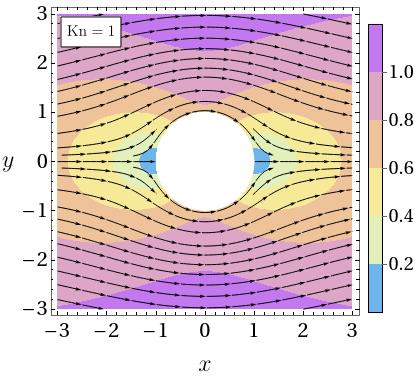} 
\caption{
\label{fig:flow_cont}Velocity streamlines over speed contours obtained from the MFS applied on the CCR model for the Knudsen numbers $\mathrm{Kn}=0.1$, $0.5$ and $1$. 
The other parameters are the same as those for Fig.~\ref{fig:speed}.}
\end{figure}

In order to validate our code, we first plot the (dimensionless) speed of the gas against the radial position (as one moves away from the cylinder) for the angles $\vartheta=0$, $\pi/4$ and $\pi/2$ in Fig.~\ref{fig:speed}. 
From left to right, the panels in the figure depict the speed of the gas for $\mathrm{Kn}=0.1$, $0.5$ and $1$. 
The solid lines in the figure delineate the results obtained from the MFS applied on the CCR model while the symbols display the results obtained from the analytic solution of the CCR model obtained in Sec.~\ref{sec:analytical}.
An excellent agreement of the results from the MFS with the analytic results---evident in the figure---validates our numerical code.
The figure reveals that the speed of the gas starts increasing for all values of $\vartheta$ as one moves away from the disk. 
For $\vartheta=0$, the speed keeps increasing with $r$ all the way to the artificial boundary. 
On the other hand, for $|\vartheta|>0$ (blue and red colors in the figure), the speed of the gas starts increasing as one moves away from the disk; the speed even surpasses its inlet value due to the accelerated flow occurring due to the production of pressure gradient around the disk; after attaining a maximum at a point somewhere in between the periphery of the disk and the artificial boundary the speed slows down on moving further away from the disk to match the fixed speed (through the boundary condition) on the artificial boundary.
The figure also shows that for $|\vartheta|>0$ (blue and red colors in the figure), the speed of the gas on the disk increases with the Knudsen number due to increasing slip velocity with the Knudsen number. 
%
%
In order to have a better idea about the speed and velocity profiles around the disk, the streamlines and speed contours obtained from the MFS results for $\mathrm{Kn}=0.1$, $0.5$ and $1$ are exhibited in Fig.~\ref{fig:flow_cont}.
While the streamlines in Fig.~\ref{fig:flow_cont} are qualitatively alike, the speed contours reveal the quantitative differences for different Knudsen numbers. 
The speed contours in Fig.~\ref{fig:flow_cont}, similarly to Fig.~\ref{fig:speed}, also show that the speed of the gas at any point in the domain increases with increasing the Knudsen number in general. 
Particularly, it is clearly visible from the speed contours in a close proximity of the disk.
Moreover, for $|\vartheta| = \pi/2$, Fig.~\ref{fig:flow_cont}---similarly to that shown by red lines in Fig.~\ref{fig:speed}---shows that the point at which the speed surpasses its inlet value of $v_0 = 1$ becomes closer and closer to the disk with increasing the Knudsen number.

%
%

\begin{figure}[!tb]
\centering
\includegraphics[scale=0.48]{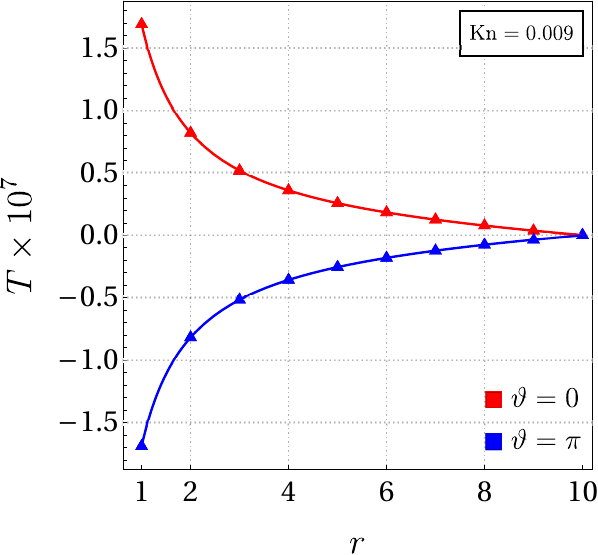}\quad \includegraphics[scale=0.46]{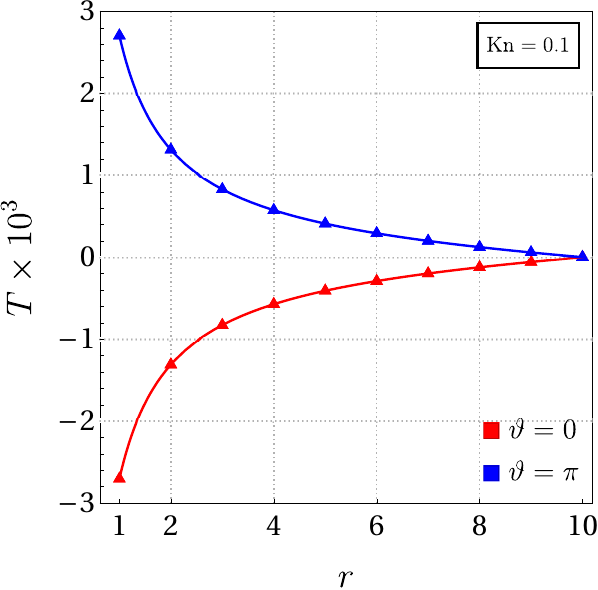} \quad
\includegraphics[scale=0.46]{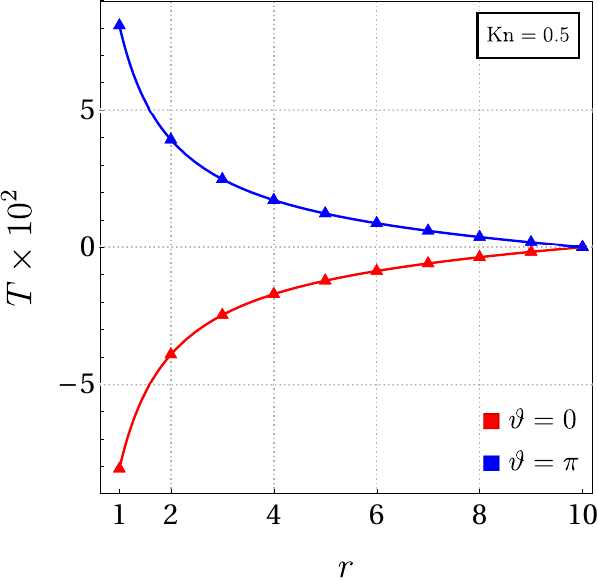} 
\caption{
\label{fig:temp_circular}Temperature along the left and right sides of the disk for $\mathrm{Kn}=0.009$, $0.1$ and $0.5$. Solid lines represent the results obtained from the MFS applied to the CCR model and the triangles represent the analytic solutions. The other parameters are the same as those for Fig.~\ref{fig:speed}.}
\end{figure}

\begin{figure}[!tb]
 \centering
 \includegraphics[scale=0.34]{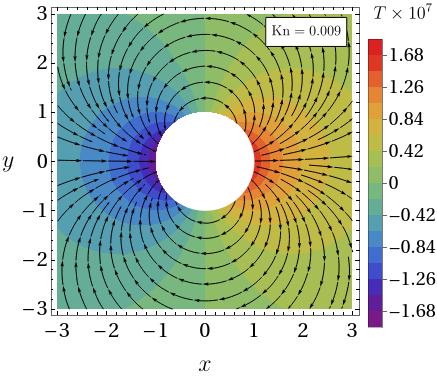}
  \includegraphics[scale=0.34]{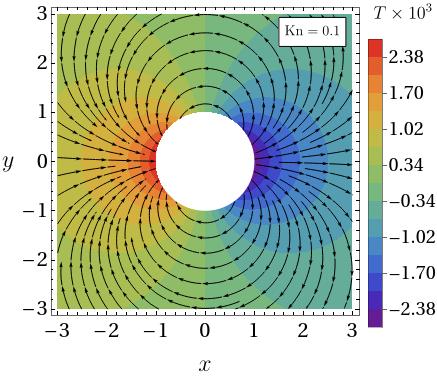}
   \includegraphics[scale=0.34]{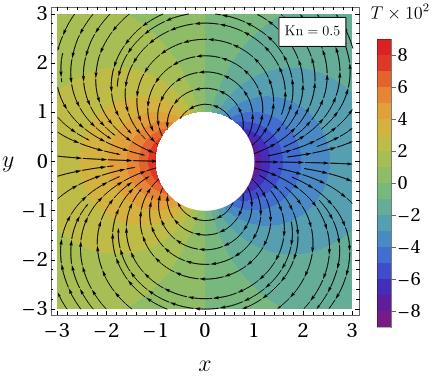}
   \caption{\label{fig:heat_circular} Heat flux lines over (scaled) temperature contours obtained from the MFS applied to the CCR model for Knudsen numbers $\mathrm{Kn}=0.009$, $0.1$ and $0.5$. The other parameters are the same as those for Fig.~\ref{fig:speed}.}
 \end{figure}

It is well established theoretically as well as experimentally that rarefied gases, when flowing around an object, manifest temperature polarization near the boundary of the object, even in the absence of any external temperature difference~\citep{BVDR1983, Torrilhon2010, RTS2013, WT2012}.
Temperature polarization is a rarefaction effect that is pertinent to rarefied gases.
To check for the temperature polarization effect in our problem, we plot the  (dimensionless) temperature of the gas at different points along the $x$-axis in Fig.~\ref{fig:temp_circular}, which illustrates the temperature on the left and right sides of the disk (i.e., along $\vartheta = \pi$ and $\vartheta=0$, respectively) for different values of the Knudsen number $\mathrm{Kn}=0.009$, $0.1$ and $0.5$.
Since the induced temperature is very small, the temperature has been scaled up by its order while depicting it  in Fig.~\ref{fig:temp_circular}.
The solid lines and symbols again denote the results obtained from the MFS applied on the CCR model and from the analytic solution, respectively, that again turn out to be in an admirable agreement.
The figure shows the presence of temperature polarization.
Nonetheless, for small Knudsen number $\mathrm{Kn}=0.009$ that corresponds to the hydrodynamic regime (the left most panel in the figure), the magnitude of temperature polarization is very small (of the order of $10^{-7}$) with minute cold and hot regions near the disk boundary along $\vartheta = \pi$ and $\vartheta=0$, respectively.
However, as the Knudsen number increases (see the middle and right panels of the figure), the magnitude of temperature polarization increases and, moreover, temperature reversal can also be seen from the middle and right panels of the figure for $\mathrm{Kn}=0.1$ and $\mathrm{Kn}=0.5$.
The temperature reversal for higher Knudsen numbers has also been seen in rarefied gas flows around spheres \cite{Torrilhon2010}. 
To get deeper insights of temperature polarization and temperature reversal, we plot the temperature contours and heat-flux lines in Fig.~\ref{fig:heat_circular}. 
The figure shows that the heat-flux lines in all panels are starting from the right side of the disk and going toward the left side of the disk for all Knudsen numbers. 
However, the temperature on the right side of the disk is higher than that on the left side only for very small Knudsen numbers (e.g.~for $\mathrm{Kn}=0.009$ in the left most panel of Fig.~\ref{fig:heat_circular}), i.e., when the flow is in the hydrodynamic regime. 
In this regime, Fourier's law remains valid and hence the heat flows from hot to cold regions. 
For large Knudsen numbers (e.g., for $\mathrm{Kn}=0.1$ and $0.5$ in the middle and right panels of Fig.~\ref{fig:heat_circular}), the temperature on the left side of the disk is higher than that on the right side due to temperature reversal and heat interestingly seems to be flowing from cold to hot regions, which is an anti-Fourier effect and is common to stress-driven rarefied gas flows; see, e.g., Refs.~\cite{Torrilhon2010, RTS2013, Gupta2015,  RGS2018, RGST2021}. 
As no temperature difference is applied externally in such problems, minuscule temperature differences are rendered by the stress gradients. 
In other words, stress gradients in such problems dominate the temperature gradients and since Fourier's law depends only on the temperature gradient, the anti-Fourier effect cannot be described by the NSF equations. 
On the other hand, the inherent coupling of the heat flux with the stress gradient in the constitutive relations of the CCR model enables it to capture the anti-Fourier effect. 
%
%
 \begin{figure}[!tb]
 \centering
 \includegraphics[scale=0.34]{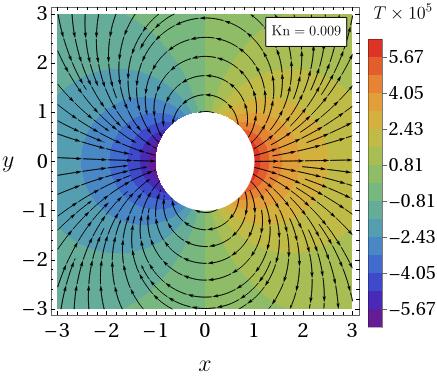}
  \includegraphics[scale=0.34]{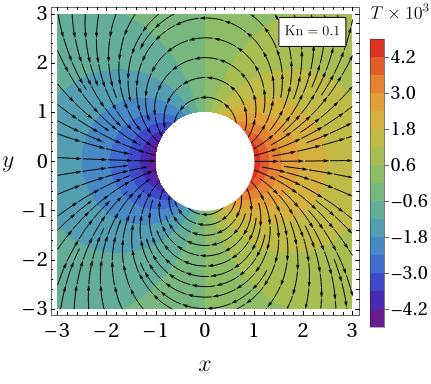}
   \includegraphics[scale=0.34]{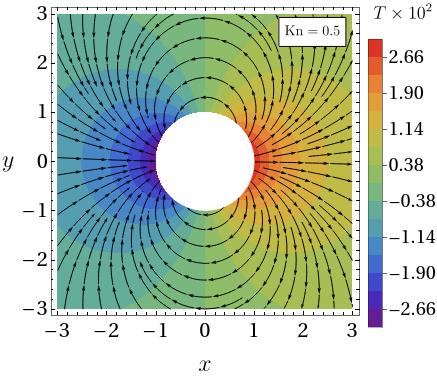}
   \caption{\label{fig:nsf_heat_circular} Heat flux lines over (scaled) temperature contours obtained from the MFS applied to the NSF model for Knudsen numbers $\mathrm{Kn}=0.009$, $0.1$ and $0.5$. The other parameters are the same as those for Fig.~\ref{fig:speed}.}
 \end{figure}
To corroborate the inability of the NSF model  in capturing the above findings,
we have also applied the MFS to the NSF model (by setting $\alpha_0=0$ in the CCR model) and displayed the temperature contours and heat-flux lines obtained from the MFS applied to the NSF model in Fig.~\ref{fig:nsf_heat_circular}. 
It turns out that the NSF model with the first-order temperature-jump boundary condition does not show temperature polarization at all (not shown here explicitly for brevity). 
With the second-order temperature jump boundary condition, the NSF model does show temperature polarization, yet reversal of temperature does not appear in order to respect imposed Fourier's law adherent to the NSF equations, which is clearly discernible in Fig.~\ref{fig:nsf_heat_circular} that has been made using the second-order velocity-slip and temperature-jump boundary conditions.
Furthermore, by comparing Figs.~\ref{fig:heat_circular} and \ref{fig:nsf_heat_circular}, it is evident that the NSF model does not show the temperature reversal. 

\begin{figure}[!tb]
\centering
\includegraphics[height=50mm]{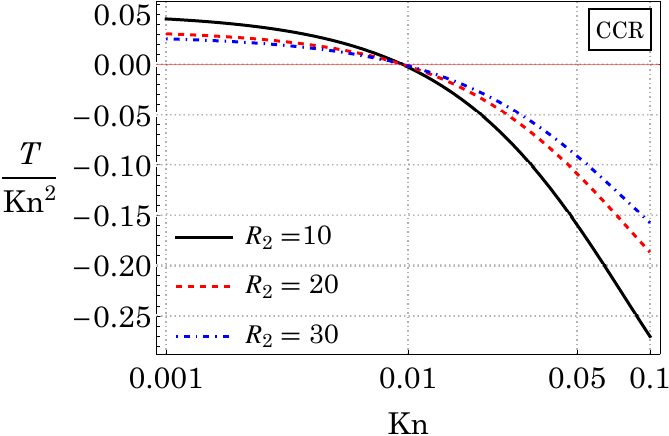} \quad
\includegraphics[height=50mm]{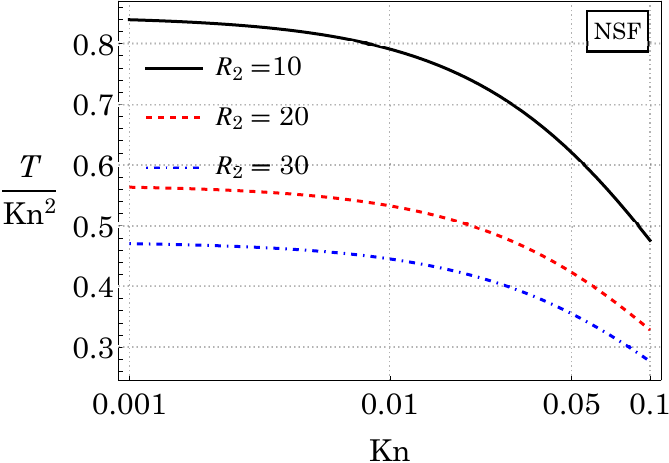} 
\caption{\label{fig:temp_reversal}
{\color{black}Temperature of the gas at $r=1$ and $\vartheta=0$ scaled with $\mathrm{Kn}^2$ plotted against the Knudsen number for different locations of the artificial boundary. 
The left panel shows the results obtained with the CCR model and the right panel shows the results obtained with the NSF equations and the second-order accurate boundary conditions.
}}
\end{figure}

{\color{black}
As also mentioned above, Fig.~\ref{fig:temp_circular} and the color bars in Figs.~\ref{fig:heat_circular} and \ref{fig:nsf_heat_circular} show that the magnitude of the temperature generated near the left and right sides of the disk increases with the increasing Knudsen number. 
Indeed, the temperature polarization and temperature reversal are second-order effects with respect to the Knudsen number. 
Hence, the generated temperature is actually of $\mathcal{O}(\mathrm{Kn}^2)$. 
In order to illustrate this, we plot the temperature of the gas at $r=1$ and $\vartheta=0$ scaled with $\mathrm{Kn}^2$ against the Knudsen number  for different locations of the artificial boundary in Fig.~\ref{fig:temp_reversal}. 
The left panel of the figure displays the results obtained with the CCR model and the right panel exhibits the results obtained with the NSF equations and the second-order accurate boundary conditions.
The figure shows that $T/\mathrm{Kn}^2$ indeed has a common scale for all Knudsen numbers.
The left panel of Fig.~\ref{fig:temp_reversal} again reveals the presence of a critical Knudsen number at which $T/\mathrm{Kn}^2$ changes its sign.
This critical Knudsen number, which is $\mathrm{Kn} \approx 0.0094115$, in the left panel of the figure demarcates the point of temperature reversal.
Evidently from the right panel, the NSF equations even with the second-order boundary conditions do not show the temperature reversal. 
Figure~\ref{fig:temp_reversal} further shows that while the temperature profiles are qualitatively similar for all locations of the artificial boundary, quantitative differences are conspicuously present. 
Indeed, the figure exhibits decreasing magnitudes of the temperature with increasing values of $R_2$.
It is interesting, however, to note that irrespective of the location of the artificial boundary, the critical Knudsen number for the temperature reversal remains fixed as evident from the left panel of Fig.~\ref{fig:temp_reversal}.
%
}

\begin{figure}[!tb]
\centering
\includegraphics[scale=0.8]{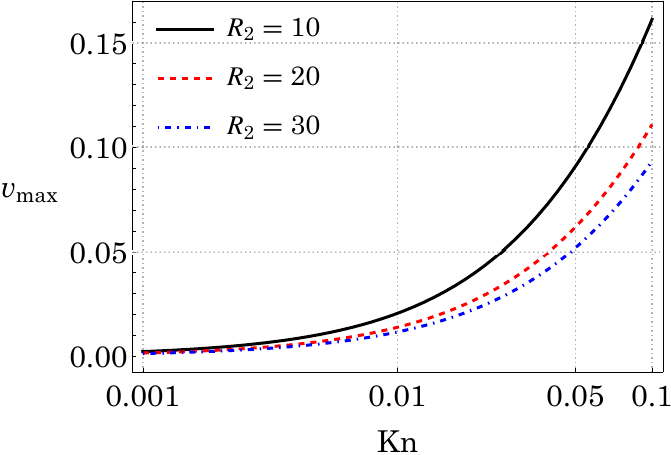} 
\caption{\label{fig:speed_max} {\color{black}Maximum speed of the gas on the disk plotted against the Knudsen number for different locations of the artificial boundary.
}
}
\end{figure}

{\color{black}
In order to check the dependence of other flow variables on the location of the artificial boundary, we also plot the maximum speed of the gas $v_{\max}=\max\{\vert\bm{v}\vert\}$ on the disk (i.e., the speed of the gas at $r=1$ and $\vartheta=\pi/2$ or $\vartheta=3\pi/2$) for different locations of the artificial boundary in Fig.~\ref{fig:speed_max}. 
Similarly to Fig.~\ref{fig:temp_reversal}, Fig.~\ref{fig:speed_max} shows that the maximum speed (i.e., the speed at $r=1$ and $\vartheta=\pi/2$ or $\vartheta=3\pi/2$) of the gas is also reduced as the distance between the artificial boundary and the actual boundary increases.
It turns out (although not shown here) that the magnitudes of the other flow variables also decrease with the increasing gap between the artificial and actual boundaries in general.
Thus, the location of the artificial boundary or the distance at which the far-field conditions are applied does influence the results quantitatively.
But, as we could not find any theoretical/numerical/experimental data on the flow variables for this problem in the literature, it is difficult to say which location of the artificial boundary gives the best results. 
Nevertheless, data on the drag force on the cylinder are available in the literature, which gives us a chance to compute the drag force on the cylinder with the method demonstrated in the present paper and to compare it with the existing results in order to decide for an appropriate location of the artificial boundary. 
Therefore, in what follows, we compute the drag force acting on the disk analytically as well as numerically through the MFS.
}

The analytic expression for the net force $\bm{F}^{(\mathrm{A})}$ acting on disk is given by the integration of the normal component of the  pressure tensor $\bm{P}(=\bm{\sigma}+p\bm{I})$ over the periphery of the disk, i.e.
\begin{align}
\label{f1}
\bm{F}^{(\mathrm{A})}=\int \bm{P}\cdot\bm{n}\, \mathrm{d}s = R_1\int_{0}^{2\pi} (\bm{P}\cdot\bm{n})\, \mathrm{d}\vartheta,
\end{align}
where $\bm{n}$ is the normal vector to the boundary and $\mathrm{d}s$ is the length of the arc that subtends angle $\mathrm{d}\vartheta$ on the center of the disk. 
The net force $\bm{F}^{(\mathrm{A})}$ can be simplified using the analytic solutions for $\bm{\sigma}$ and $p$ determined in Sec.~\ref{sec:analytical}.
The drag force on the disk is given by the projection of the net force in the upstream direction, i.e.~by
\begin{align}
F_d^{(\mathrm{A})} = - \bm{F}^{(\mathrm{A})} \cdot \hat{\bm{x}},
\end{align}
where $\hat{\bm{x}}$ denotes the unit vector in the downstream direction. 
On simplification the (analytic) drag force turns out to be
\begin{align}
\label{drag_analytic}
F_d^{(\mathrm{A})} = 4\pi \mathrm{Kn} c_6,
\end{align}
where the value of the constant $c_6$ is evaluated from the boundary conditions and hence changes with the values of the parameters $\mathrm{Kn}$ and $\alpha_0$. 
In order to calculate the net force acting on the disk through the MFS, all the point force vectors acting on the singularity points lying on the inner fictitious boundary inside the disk are superimposed, i.e. 
\begin{align}
\bm{F}^{\mathrm{(MFS)}}
=\sum_{i=1}^{N_{s_1}}\bm{f}^{(i)}.
\end{align}
The drag on the disk is again given by the projection of the net force in the upstream direction, i.e.~by
\begin{align}
F_d^{\mathrm{(MFS)}}=-\sum_{i=1}^{N_{s_1}}\bm{f}^{(i)}\cdot \hat{\bm{x}}=-\sum_{i=1}^{N_{s_1}}f_1^{(i)}.
\end{align}
%
%
\begin{figure}[!tb]
\centering
\includegraphics[scale=0.8]{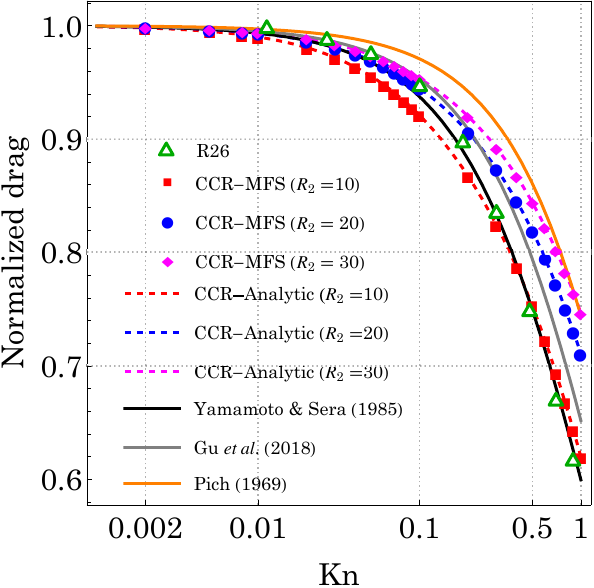} 
\color{black}{\caption{\label{fig:circular_drag}
Normalized drag on the cylinder plotted against the Knudsen number for different locations of the artificial boundary.
The dashed red, blue and magenta lines represent the analytic solution of the CCR model for $R_2=10$, $20$ and $30$, respectively. 
The square (red), disk (blue) and diamond (magenta) symbols represent the numerical solution of the CCR model obtained with the MFS for $R_2=10$, $20$ and $30$, respectively. 
The solid orange, black and gray   lines depict the normalized drag obtained with the analytic expressions of \citet{Pich1969}, \citet{YS1985} and \citet{GBJE2019}.
The green triangle symbol shows the normalized drag computed with the nonlinear R26 equations in Ref.~\cite{GBJE2019}.
The other parameters are the same as those for Fig.~\ref{fig:speed}.}}
\end{figure}
{\color{black}For illustrative purpose, it is convenient to compare the normalized drag, defined by the drag force  normalized with the Stokes drag (drag force in the limit $\mathrm{Kn}\to 0$). 
In the following, we shall investigate the effect of the location of the artificial boundary on the normalized drag. 
But, prior to this, the work of \citet{GBJE2019} must be acknowledged wherein the authors performed a thorough study of the drag coefficient for the problem of flow past a circular cylinder and gave analytic expressions for the drag coefficient valid in the continuum, slip and transition regime.
In addition, they also computed the drag coefficient for the problem through the nonlinear R26 equations.
The drag coefficient on dividing by its value in the limit $\mathrm{Kn}\to 0$ is exactly the same as the normalized drag.
This gives us an opportunity to compare the normalized drag obtained in the present paper with that obtained using the results presented in Ref.~\cite{GBJE2019}. 
Figure~\ref{fig:circular_drag} illustrates the variation in the normalized drag with the Knudsen number on changing the location of the artificial boundary.
The dashed red, blue and magenta lines in the figure depict the analytic solution of the CCR model for $R_2=10$, $20$ and $30$, respectively, while the square (red), disk (blue) and diamond (magenta) symbols denote the solution obtained from the MFS applied on the CCR model for $R_2=10$, $20$ and $30$, respectively.
The solid orange, black and gray   lines in the figure delineate the normalized drag obtained with the analytic expressions of \citet{Pich1969}, \citet{YS1985} and \citet{GBJE2019} at the Reynolds number $\mathrm{Re}=0.5$. 
The green triangle symbol depicts the  normalized drag computed directly from the data on the drag coefficient that have been obtained with the nonlinear R26 equation in Ref.~\cite{GBJE2019}.
The figure shows that the normalized drag obtained numerically with the MFS applied on the CCR model is in excellent agreement with that obtained with the analytic solution of the CCR model, irrespective of the location of the artificial boundary, which is no surprise as both (analytic and numerical) methods use a common location of the artificial boundary. 
Notwithstanding, these results do validate the correctness of our MFS-based numerical framework one more time in spite of the fact that the location of the artificial boundary does affect the results.
The figure also reveals that while the normalized drag obtained from the CCR model in the present paper is in qualitatively good agreement with the normalized drag obtained from other methods existing in the literature, quantitative differences are certainly there. 
Consequently, it is hard to tell a universal location of the artificial boundary (or, in other words, a fixed value of $R_2$) that can lead to the best results for all quantities.
However, as the normalized drag predicted by the CCR model for $R_2=10$ is generally close to that predicted by \citet{YS1985} and to the R26 data taken from Ref.~\cite{GBJE2019}, we take $R_2=10$ throughout the paper.

}

{\color{black} 
Although we have presented the results for somewhat large Knudsen numbers and the normalized drag---being a global quantity---also turned out to be agreeing well with that obtained with other models, it is important to note that the flow profiles predicted by the CCR model are accurate only in the bulk and only for relatively small Knudsen numbers ($\mathrm{Kn} \lesssim 0.2$) and that the flow profiles predicted by the CCR model near the boundary of the cylinder may differ from the actual kinetic data, especially for large Knudsen numbers. 
The reason for this is that Knudsen layers become more and more prominent near the boundary of the disk with the increasing Knudsen number and the CCR model cannot describe them.
This is a limitation of the CCR model and better continuum models, e.g.~the R13 or R26 equations, are needed to obtain an accurate flow description near the boundary of the disk. 
The R13 and R26 equations can be expected to give correct flow profiles for the Knudsen number up to $0.5$ and $1$, respectively.
Notwithstanding, the fundamental solutions of the R13 equations are not available in two dimensions and that of the R26 equations are available in any dimensions at present, and exploring them is beyond the scope of the present paper.}


\section{\label{sec:half}Results and discussion for flow past a semicircular cylinder}
The numerical framework developed in the present paper can be employed to investigate other quasi-two-dimensional flow problems as well. 
In particular, the expediency  of the method is notable for problems wherein either an analytic solution cannot be found or is arduous to find.%

\begin{figure}[!tb]
\centering
\includegraphics[scale=0.8]{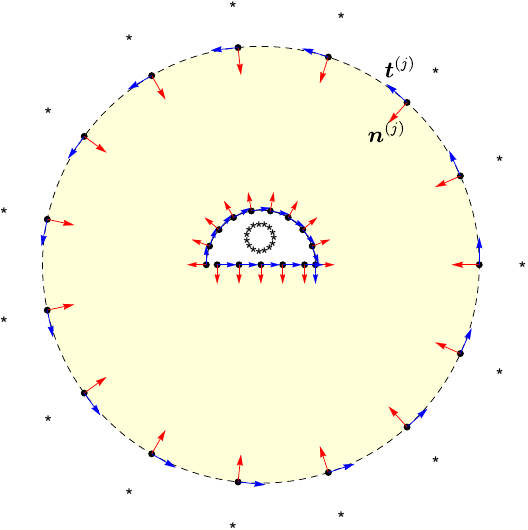}
\caption{\label{fig:half_schematic}
Schematic representation for arrangement of singularities (stars) and boundary nodes (dots). The red and blue arrows represent the normal and tangent vectors at each boundary node.} 
\end{figure}

To showcase the capabilities of the method, we now consider a problem, where the radial symmetry is absent. 
We consider the problem of rarefied gas flow past an infinitely long semicircular cylinder in its transverse direction.
The problem is still quasi-two-dimensional but flow behavior changes according to the orientation of the cylinder. 
%
\begin{figure}[!b]
\centering
\includegraphics[scale=0.8]{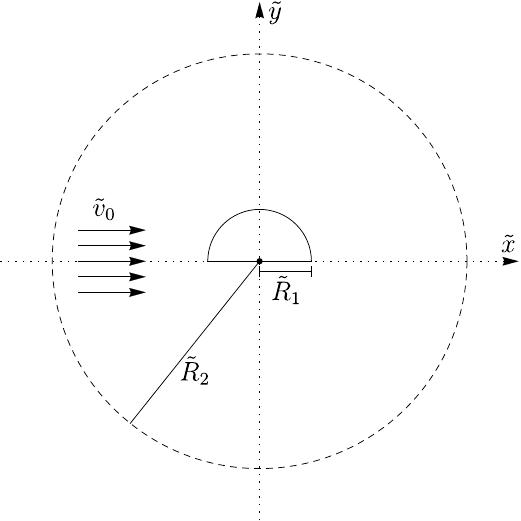} \quad
\includegraphics[scale=0.8]{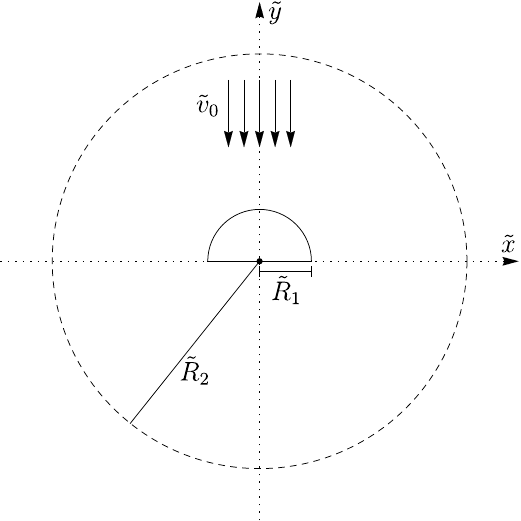} 
\caption{\label{fig:half_disk}Cross-sectional view of horizontal and vertical flow past a semicircular cylinder.}
\end{figure}
To setup the orientation of the cylinder and the flow direction, let an infinitely long semicircular cylinder of radius $\tilde{R}_1$ be placed in such a way that its axis is along the $\tilde{z}$-direction and its semicircular base is in the upper half of the $\tilde{x}\tilde{y}$-plane with the midpoint of the diameter of the semicircular base being fixed at the origin of the Cartesian coordinate system $(\tilde{x},\tilde{y},\tilde{z})$ as shown in Fig.~\ref{fig:half_disk}.
Two cases are considered: (i) a rarefied monatomic gas approaching the cylinder from the negative $\tilde{x}$-direction; we refer to this case as the case of horizontal flow or simply the horizontal case, and 
(ii) a rarefied monatomic gas approaching the cylinder from the positive $\tilde{y}$-direction; we refer to this case as the case of vertical flow or simply the vertical case analogously. 
A schematic exhibiting the cross-sectional view of both cases has also been shown in Fig.~\ref{fig:half_disk}. 
Furthermore, it is assumed that the temperature at the surface of the cylinder is the same as the far-field ambient temperature of the gas $\tilde{T}_0$.
Needless to say, we shall solve the problem in the $\tilde{x}\tilde{y}$-plane or equivalently in the $\tilde{r}\vartheta$-plane, where $\tilde{x} = \tilde{r}\cos{\vartheta}$ and $\tilde{y} = \tilde{r}\sin{\vartheta}$, for the semicircular disk in the dimensionless form.
The radius of the disk $\tilde{R}_1$ is taken as the characteristic length scale $\tilde{L}$ for non-dimensionalization so that $x = \tilde{x} / \tilde{R}_1$, $y = \tilde{y} / \tilde{R}_1$, $r = \tilde{r} / \tilde{R}_1$, and the dimensionless radius of the disk $R_1=\tilde{R}_1/\tilde{L}=1$. 
To circumvent Stokes' paradox, we---similarly to the above---place an artificial circular boundary of radius $\tilde{R}_2$ centered at $(0,0)$ sufficiently far from the semicircular disk. 
The dimensionless radius of the artificial boundary is $R_2=\tilde{R}_2/\tilde{L}$, where $R_2 > R_1$.
Furthermore, for implementation of the MFS, we also introduce two fictitious boundaries, one inside the semicircular disk and other outside the artificial circular boundary, on which the source points are to be placed. 
Let the inner fictitious boundary be a circle of radius $\tilde{R}_1^\prime$ centered at $(0,0.5)$ and the outer fictitious boundary be a circle of radius $\tilde{R}_2^\prime$ centered at $(0,0)$.
The dimensionless radii of the inner and outer fictitious boundaries are $R_1^\prime = \tilde{R}_1^\prime / \tilde{L}$ and $R_2^\prime = \tilde{R}_2^\prime / \tilde{L}$.
An illustration exhibiting the boundary nodes on the semicircular disk and on the artificial boundary, and the location of source points on the fictitious boundaries is shown in Fig.~\ref{fig:half_schematic}.
Once the singularities are placed, the rest of the procedure of implementing the MFS remains the same as explained in Sec.~\ref{sec:implement}.


The horizontal and vertical flow cases are covered by changing the boundary conditions on the artificial boundary.
For the horizontal case, the boundary conditions on the artificial boundary are
\begin{align}
\label{bc_hor}
v_x=v_0,\quad v_y=0,\quad T=0, 
\end{align}
while for the vertical case, the boundary conditions on the artificial boundary are
\begin{align}
\label{bc_ver}
v_x=0,\quad v_y=-v_0,\quad T=0.
\end{align}
The boundary conditions on the actual periphery of the disk remain the same as  boundary conditions \eqref{bc_1}--\eqref{bc_3}.
%
%

In numerical computations, $v_0$ is taken as unity, the number of boundary nodes on the actual periphery of the disk is taken as $N_{b_1}=200$ and that on the artificial boundary is taken as $N_{b_2}=400$, and the number of singularity points on the inner and outer fictitious boundaries are taken as $N_{s_1}=200$ and $N_{s_2}=400$, respectively. 
The dimensionless radius of the semicircular disk is $R_1=1$ and the dimensionless radius of the artificial boundary is taken as $R_2=10$. 
{\color{black}Although the dependence of the results on the location of the artificial boundary cannot be neglected, the validation of the results done using $R_2=10$ in Sec.~\ref{sec:results1} suggests the sufficiency for fixing $R_2$ to $10$ to get adequate qualitative results for the current problem as well.} 
The dimensionless radii of the inner and outer fictitious boundaries are taken as $R_1^\prime=0.1$ and $R_2^\prime=50$, respectively.
In the case of a circular cylinder, we had the advantage of having an analytic solution, allowing us to validate our results even with relatively fewer numbers of boundary nodes and singularity points ($N_{b_1}=N_{s_1}=50$ and $N_{b_2}=N_{s_2}=100$). 
However, when dealing with the case of a semicircular cylinder, we have taken a relatively larger number of boundary nodes and singularities. 
This decision is based on the studies from the existing literature~\citep{CH2020,FR2002}, which suggest that more boundary nodes and singularity points in the method of fundamental solution lead to improved accuracy. 
Furthermore, as it is neither easy to obtain an analytic solution for the present problem nor could we find any experimental or theoretical study in the existing literature,
our focus remains only on the qualitative analysis of the results.


\subsection{\label{sec:horizontal}Results in the case of horizontal flow}
\begin{figure}[!tb]
\centering 
\includegraphics[scale=0.35]{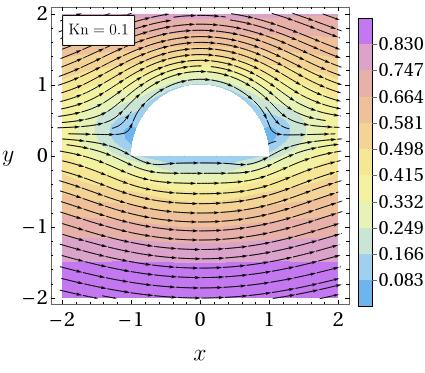} 
\includegraphics[scale=0.35]{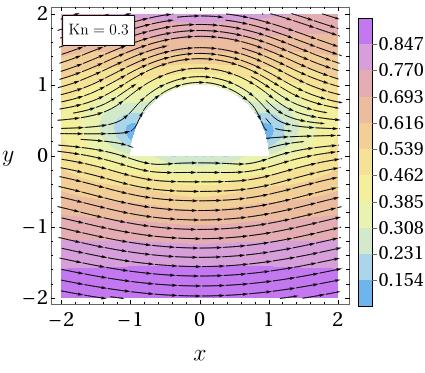} 
\includegraphics[scale=0.35]{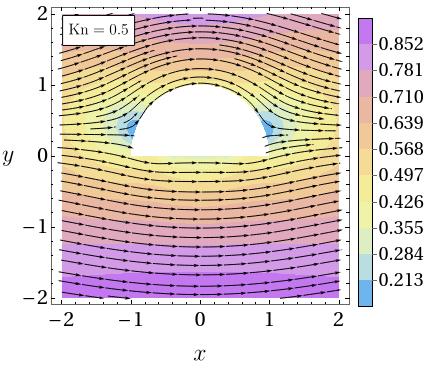} 
\caption{\label{fig:horz_stream}
Velocity streamlines along with contour plots of the speed in the background obtained from the MFS applied on the CCR model for $\mathrm{Kn}=0.1$, $0.3$ and $0.5$. 
The other parameters are $R_1=1$, $R_2=10, R_1^\prime=0.1$, $R_2^\prime=50$, $N_{b_1}=N_{s_1}=200$ and $N_{b_2}=N_{s_2}=400$.}
\end{figure}

Figure~\ref{fig:horz_stream} illustrates the velocity streamlines around the semicircular disk along with density plots of the speed in the background for $\mathrm{Kn}=0.1$, $0.3$ and $0.5$ in the case of horizontal flow, i.e.~when the flow is along the $x$-direction. 
Analogously to the problem of flow past a circular cylinder demonstrated in Sec.~\ref{sec:results1}, the streamlines in Fig.~\ref{fig:horz_stream} are qualitatively alike for the considered Knudsen numbers. 
Nonetheless, contour plots of the speed do depict quantitative differences in the speed of the gas for different Knudsen numbers that are prominently discernible in the close proximity of the disk.
It is evident from the colors of the contour plots near the disk that the speed of the gas on the disk increases with the Knudsen number due to increase in the slip velocity with an increasing Knudsen number.
%
Apparently, it is true even for any point in the domain that the speed of the gas at this point increases with increasing the Knudsen number.


\begin{figure}[!htb]
\centering 
\includegraphics[scale=0.35]{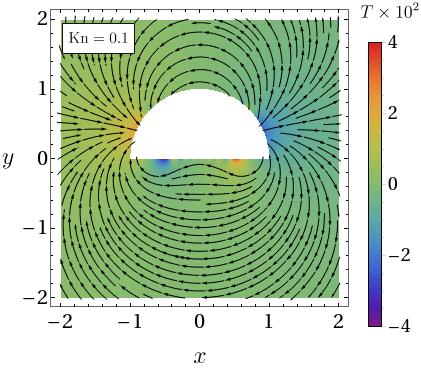} 
\includegraphics[scale=0.35]{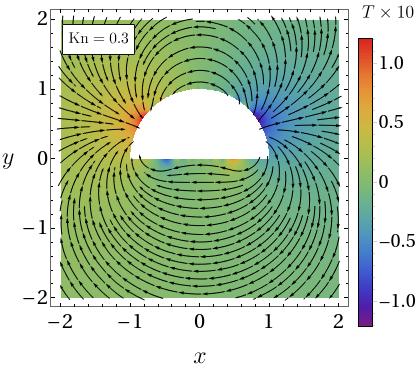} 
\includegraphics[scale=0.35]{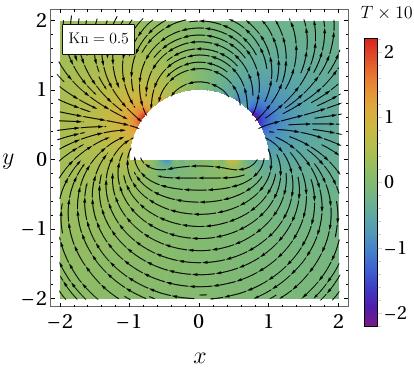} 
\caption{\label{fig:horz_heat}
Heat flux lines along with density plots of the temperature in the background obtained from the MFS applied on the CCR model for $\mathrm{Kn}=0.1$, $0.3$ and $0.5$. 
The other parameters are the same as those for Fig.~\ref{fig:horz_stream}.}
\end{figure}

Interestingly, the effects of asymmetry in the shape of the object are revealed when the variation of temperature of the gas is explored.
In order to explore the asymmetry effects, we plot in Fig.~\ref{fig:horz_heat} the heat flux lines superposed on density plots of the temperature for $\mathrm{Kn}=0.1$, $0.3$ and $0.5$. 
The figure reveals the existence of temperature polarization near the disk  for all Knudsen numbers---with hot region (denoted by red color) on the left side of the curved portion of the disk and cold region (denoted by blue color) on the right side due to compression (expansion) of the gas on the left (right) side.
In addition, a minute (but opposite in sign) temperature polarization also occurs below the flat portion of the disk and is conspicuous for small Knudsen numbers (for $\mathrm{Kn}=0.1$ in the figure) but diminishes as the Knudsen number increases.
This double polarization could be attributed to the presence of corners in the geometry or to the asymmetry present in the geometry.
As the Knudsen number increases, the strength of temperature polarization on the curved portion of the disk increases and hence it takes over the minute temperature polarization below the flat portion of the disk, and the latter fades away gradually as the Knudsen number increases.
The heat flux lines in Fig.~\ref{fig:horz_heat} indicate the flow of heat from cold to hot regions, depicting the anti-Fourier effect that again cannot be captured with the classical models in fluid dynamics.


\subsection{Results in the case of vertical flow}
\begin{figure}[!tb]
\centering
\includegraphics[scale=0.35]{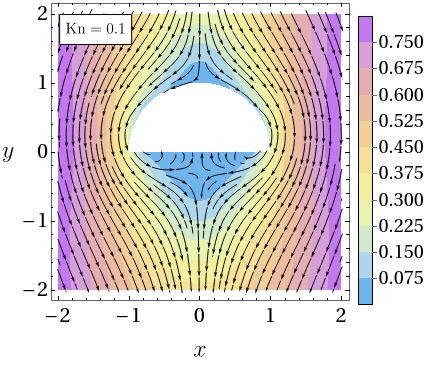} 
\includegraphics[scale=0.35]{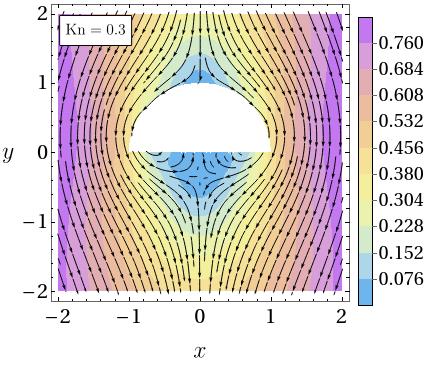} 
\includegraphics[scale=0.35]{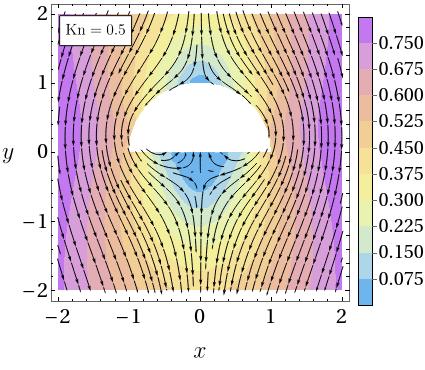} 
\caption{\label{fig:vert_stream}Velocity streamlines along with contour plots of the speed in the background obtained from the MFS applied on the CCR model for $\mathrm{Kn}=0.1$, $0.3$ and $0.5$.
The other parameters are the same as those for Fig.~\ref{fig:horz_stream}.}
\end{figure}

Figure~\ref{fig:vert_stream} exhibits the velocity streamlines around the semicircular disk along with stream plots of the speed in the background for $\mathrm{Kn}=0.1$, $0.3$ and $0.5$ in the case of vertical flow, i.e.~when the flow is along the negative $y$-direction.
The figure presents flow separation and formation of circulation zones after the flow crosses the disk.
The figure shows that the flow separation starts reducing slightly with increasing the Knudsen number.
Flow separation and an analogous outcome---reduction in the size circulation zone with decreasing Reynolds number---have also been reported by \citet{NL2020} for a creeping (or low-Reynolds-number) flow past a semicircular cylinder.
Thus, owing to the inverse relationship between the Reynolds number and the Knudsen number, the qualitative nature of the flow predicted by the CCR model in the present paper is justified.
Contour plots of the speed in Fig.~\ref{fig:vert_stream} again depict that the speed of the gas around the disk increases with increase in the Knudsen number.
%
%
%
\begin{figure}[!tb]
\centering
\includegraphics[scale=0.35]{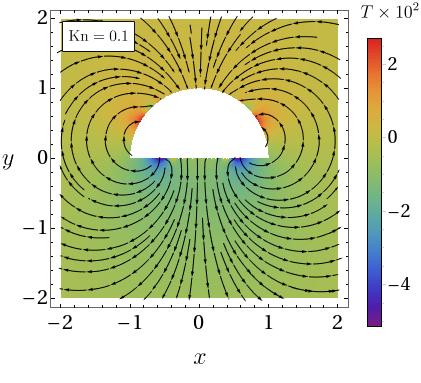} 
\includegraphics[scale=0.35]{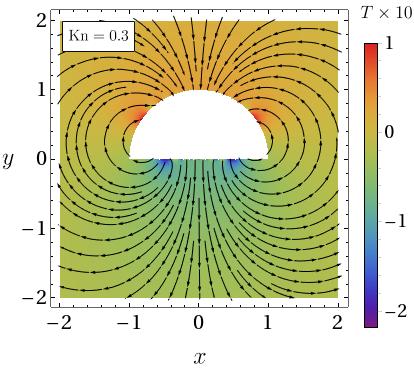} 
\includegraphics[scale=0.35 ]{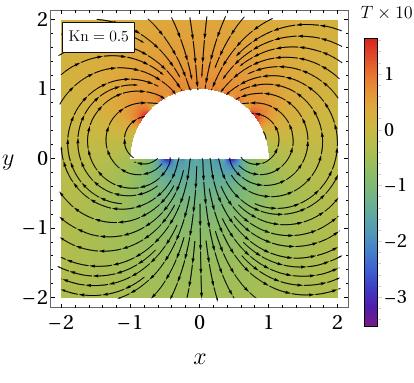} 
\caption{\label{fig:vert_heat}Heat flux lines along with density plots of the speed in the background obtained from the MFS applied on the CCR model for $\mathrm{Kn}=0.1$, $0.3$ and $0.5$. The other parameters are the same as those for Fig.~\ref{fig:horz_stream}.}
\end{figure}

Figure \ref{fig:vert_heat} illustrates the heat flux lines superposed over density plots of the temperature for $\mathrm{Kn}=0.1$, $0.3$ and $0.5$ in the case of vertical flow. 
Temperature polarization again occurs in this case but it is symmetric about the $y$-axis in this case, creating hot and cold regions on the top and bottom of the disk, respectively.
The strength of temperature polarization increases with increase in the Knudsen number.
The heat flux lines are also symmetric about the $y$-axis for all Knudsen numbers and show the heat flowing from cold to hot regions, illustrating the anti-Fourier effect in the present case as well. 

\subsection{Drag force in the horizontal and vertical cases}
\begin{figure}[!htb]
\centering
\includegraphics[scale=0.7]{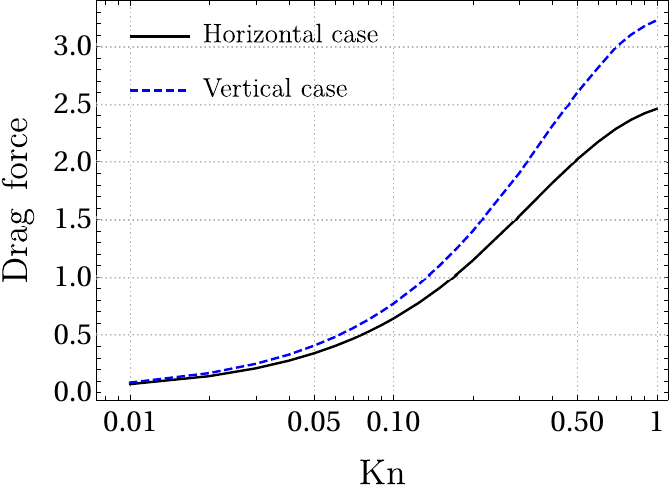} 
\caption{\label{fig:drag_half}Drag force on the semicircular disk plotted against the Knudsen number in the horizontal and vertical cases. 
The other parameters are the same as those for Fig.~\ref{fig:horz_stream}.}
\end{figure}

To the best of our knowledge, an analytic expression or any experimental result for the drag force exerted on the semicircular disk in this problem does not exist in the literature. 
Therefore, we directly present the drag force predicted by the CCR model through the MFS  in Fig.~\ref{fig:drag_half} for the horizontal and vertical cases. 
The drag force in the horizontal case has been obtained by taking the projection of net force in the negative $x$-direction (similarly to that in the problem of flow past a circular cylinder in Sec.~\ref{sec:results1}) and is demarcated in Fig.~\ref{fig:drag_half} by the solid (black) line while the drag force in the vertical case has been obtained by taking the projection of net force in the positive $y$-direction and is demarcated in Fig.~\ref{fig:drag_half} by the dashed (blue) line.
Similarly to the drag force on the circular cylinder obtained in Sec.~\ref{sec:results1}, Fig.~\ref{fig:drag_half} shows that the drag force increases with increasing the Knudsen number in both horizontal and vertical cases. 
However, unlike the case of a circular cylinder where the dependence of the drag force on the Knudsen number was apparent through Eq.~\eqref{drag_analytic}, an expression revealing dependence of the drag force on the Knudsen number in the case of a semicircular cylinder is lacking unfortunately.



\section{\label{sec:conclusion}Conclusion}
Slow transverse-directional flows of a rarefied monatomic gas past a circular cylinder and past a semicircular cylinder have been investigated with the CCR model.
The CCR model, being a refined model than the classical NSF equations, has allowed us to study moderately rarefied gas flows by means of the MFS, an efficient meshfree numerical method. 
Owing to the uniformity of the flow along the axes of the cylinders, the problem of flow past a right circular cylinder reduces to the problem of flow past a circular disk and the problem of flow past a right semicircular cylinder to the problem of flow past a semicircular disk; and hence both problems have essentially been investigated in two dimensions. 
{\color{black}To address the limitations imposed by Stokes' paradox in studying flow past two-dimensional objects, the domain has been made ``bounded" artificially by introducing an artificial boundary in the flow domain far from the disk. 
Appropriate boundary conditions have been imposed on this  artificial boundary ensuring that the nature of the flow remains consistent despite the presence of the artificial
boundary. 
This has allowed us to obtain a meaningful analytic solution for the flow past a circular disk and to use it to validate our numerical framework based on the MFS.
It is, however, worth mentioning that the obtained solutions (analytic as well as that obtained through the MFS) do depend on the location of the artificial boundary. 
To eliminate the dependency of the solution on the location of the artificial boundary, the only apparent way is to not use the artificial boundary but get rid of with Stokes' paradox by some other means, e.g., by including convective terms in the CCR model (like Oseen's correction to the Stokes equations).
Inclusion of convective terms in the CCR model would, however, require deriving and implementing the corresponding fundamental solutions, which is beyond the scope of the present paper and will be explored elsewhere in the future. 
}

The numerical results for the physical quantities, such as the velocity, temperature and drag force, obtained from the MFS applied on the CCR model have shown an excellent agreement with the analytic solution of the CCR model. 
{\color{black}The normalized drag obtained from the MFS applied on the CCR model in the present paper agrees quite well with the analytic solution and reasonably well even with the results on the normalized drag available in the literature
~\citep{GBJE2019,YS1985,Pich1969}.}
It has been found that the CCR model is able to capture the well-known rarefaction effects pertaining to the problem of flow past a circular disk, e.g., temperature polarization, temperature reversal and, especially, the anti-Fourier heat transfer and that the last two effects could not be captured by the NSF equations.
To demonstrate the capabilities of the developed numerical framework, the problem of flow past a semicircular disk has also been investigated for which an analytic solution of any model (even the Stokes equations) does not exist as per our knowledge.
The temperature polarization and anti-Fourier heat transfer have been revealed by the CCR model for this problem as well.
Overall, the results obtained from the MFS applied to the CCR model provided valuable insights into rarefied gas flows past circular and semicircular cylinders. 
The ultimate usefulness of the present paper will be revealed once it can be used to investigate the problem of flow past an object of an arbitrary shape.
For this, although there is nothing special required to be done with the developed numerical framework, defining the boundary of the object mathematically and hence defining the normal and tangent vectors at the boundary nodes require some effort.
So the problem of rarefied gas flow past objects of arbitrary shapes will be a subject for future research.
Notwithstanding, the present paper does provide a direction toward this goal.

\section*{Acknowledgements}
Himanshi gratefully acknowledges the financial support from the Council of Scientific and Industrial Research (CSIR) [File No.: 09/1022(0111)/2020-EMR-I].  A.S.R. acknowledges the~financial support from the Science and Engineering Research Board, India through the Grant No. MTR/2021/000417. Himanshi and V.K.G. also acknowledge the facilities of the Bhaskaracharya Mathematics Laboratory and Brahmagupta Mathematics Library supported by DST-FIST Project SR/FST/MS I/2018/26 that have been used to carry out this work.\\

\appendix
\numberwithin{equation}{section}
\section{\label{app:A}Fundamental solutions of the CCR model}
In order to find the fundamental solutions, we use the Fourier transformation which is defined as
\begin{align}
\label{ft}
\mathcal{F}\big(F(\bm{r})\big) 
=\hat{F}(\bm{k}) := \int_{\mathbb{R}^2} F(\bm{r}) \, \mathrm{e}^{\mathbbm{i} \, \bm{k} \cdot \bm{r}} \, \mathrm{d} \bm{r}
\end{align}
and the corresponding inverse Fourier transformation is defined as
\begin{align}
\label{ftinv}
\mathcal{F}^{-1}\big(\hat{F}(\bm{k})\big) 
=F(\bm{r})
:=\frac{1}{(2\pi)^2}\displaystyle{ \int_{\mathbb{R}^2}}
\hat{F}(\bm{k}) \, \mathrm{e}^{-\mathbbm{i}\,\bm{k} \cdot \bm{r}} \, \mathrm{d}\bm{k},
\end{align}
where $\bm{k}$ is the wavevector in the spatial-frequency domain and $\mathbbm{i}$ is the imaginary unit. 
Now, we derive the fundamental solutions (previously derived in Ref.~\citep{HRG2023}) via alternate approach of considering two different cases by incorporating the sourcing terms separately in the momentum balance and the energy balance equation.
In the first case, a sourcing term is considered in the momentum balance equation, which is a point force vector $f_i$. The balance equations~\eqref{cons_laws} in indicial notations read
\begin{align}
\label{mass_bl}
\frac{\partial v_i}{\partial x_i}&=0,
\\
\label{mom_bl}
\frac{\partial p}{\partial x_i}+\frac{\partial \sigma_{ij}}{\partial x_j}&=f_i \,\delta(\bm{r}),
\\
\label{energy_bl}
\frac{\partial q_i}{\partial x_i}&=0,
\end{align}
where $\bm{r}= (x_1,x_2)^{\mathsf{T}}$.
The constitutive relations~\eqref{ccr} read
\begin{align}
\label{ccr_1_q2D}
\sigma_{ij}=&-2\mathrm{Kn}\left[\frac{1}{2}\left(\frac{\partial v_i}{\partial x_j}+\frac{\partial v_j}{\partial x_i}\right) - \frac{1}{3} \delta_{ij} \frac{\partial v_\ell}{\partial x_\ell}\right]
-2\alpha_0\mathrm{Kn}\left[\frac{1}{2}\left(\frac{\partial q_i}{\partial x_j}+\frac{\partial q_j}{\partial x_i}\right) - \frac{1}{3} \delta_{ij} \frac{\partial q_\ell}{\partial x_\ell}\right],
\\
\label{ccr_2_q2D}
q_i=&-\frac{c_p \mathrm{Kn}}{\mathrm{Pr}}\left(\frac{\partial T}{\partial x_i}+\alpha_0 \frac{\partial \sigma_{ij}}{\partial x_j}\right).
\end{align}
Applying the Fourier transformation in Eqs.~\eqref{mass_bl}--\eqref{energy_bl}, \eqref{ccr_1_q2D} and \eqref{ccr_2_q2D} and using the fact that $\mathcal{F}[\delta(\bm{r})]=1$,  we obtain ($i,j,\ell \in \{1,2\}$)
\begin{align}
\label{mass_ft}
k_i \hat{v}_i &=0,
\\
\label{mom_ft}
k_i \hat{p} + k_j \hat{\sigma}_{ij} &= \mathbbm{i} \, f_i,
\\
\label{energy_ft}
k_i \hat{q}_i &= 0,
\\
\label{ccr1_ft}
\hat{\sigma}_{ij}&=\medspace \mathbbm{i} \,\mathrm{Kn} \bigg[ k_j(\hat{v}_i + \alpha_0\hat{q}_i)+k_i(\hat{v}_j+\alpha_0\hat{q}_j)-\frac{2}{3} \delta_{ij} k_\ell (\hat{v}_\ell+\alpha_0\hat{q}_\ell)\bigg],
\\
\label{ccr2_ft}
\hat{q}_i &=\medspace \mathbbm{i}\frac{c_p \mathrm{Kn}}{\mathrm{Pr}}\left(k_i \hat{T}+\alpha_0 k_j\hat{\sigma}_{ij}\right),
\end{align}
where the variables with hat are the Fourier transforms of the corresponding field variables. 
Using Eqs.~\eqref{mass_ft} and \eqref{energy_ft},  Eq.~\eqref{ccr1_ft} simplifies to
\begin{align}
\label{ss1}
\hat{\sigma}_{ij}=& \medspace\mathbbm{i} \,\mathrm{Kn} \big[ k_j(\hat{v}_i + \alpha_0\hat{q}_i)+k_i(\hat{v}_j+\alpha_0\hat{q}_j)\big].
\end{align}
Multiplying the above equation with  $k_j$ and $k_i k_j$, we obtain \begin{align}
\label{ss11}
k_j\hat{\sigma}_{ij}&=\mathbbm{i} \, \mathrm{Kn}\, k^2(\hat{v}_i+\alpha_0\hat{q}_i),
\\
\label{ss12}
k_i k_j\hat{\sigma}_{ij}&=0,
\end{align}
respectively, where $k_i k_i=|k_i|^2=k^2$ has been used.
Multiplying Eq.~\eqref{ccr2_ft} with $k_i$ and utilizing Eqs.~\eqref{energy_ft} and \eqref{ss12},  we obtain 
\begin{align}
\label{theta_ft}
\hat{T}=0.
\end{align}
Again, multiplying Eq.~\eqref{mom_ft} with $k_i$ and utilizing Eq.~\eqref{ss12},  we obtain 
\begin{align}
\label{p_ft}
\hat{p}&=\mathbbm{i}\frac{k_i f_i}{k^2}.
\end{align}
Now, from Eqs.~\eqref{mom_ft} and \eqref{p_ft}, one can easily write \begin{align}
\label{ss5}
k_j\hat{\sigma}_{ij}&=\mathbbm{i} f_i-\mathbbm{i}\frac{k_i k_j f_j}{k^2}.
\end{align}
Substituting the value of $\hat{T}$ from Eq.~\eqref{theta_ft} and the value of $k_j\hat{\sigma}_{ij}$ from Eq.~\eqref{ss5} into Eq.~\eqref{ccr2_ft}, we obtain
\begin{align}
\label{q_ft}
\hat{q}_i&= -\frac{c_p \mathrm{Kn}}{\mathrm{Pr}} \alpha_0 f_j\left(\delta_{ij}-\frac{k_ik_j}{k^2}\right).
\end{align}
Now, from Eqs.~\eqref{ss11}, \eqref{ss5} and \eqref{q_ft},
\begin{align}
\label{v_ft}
\hat{v}_i=& \medspace \frac{f_j}{\mathrm{Kn}}\left(\frac{\delta_{ij}}{k^2}-\frac{k_i k_j}{k^4}\right)+\frac{c_p \mathrm{Kn}}{\mathrm{Pr}} \alpha_0^2 f_j\left(\delta_{ij}-\frac{k_i k_j}{k^2}\right).
\end{align}
Finally,  using Eqs.~\eqref{q_ft} and \eqref{v_ft} in Eq.~\eqref{ccr1_ft}, we obtain \begin{align}
\label{sigma_ft}
\hat{\sigma}_{ij}=&\medspace \mathbbm{i} \, f_\ell 
\left(\frac{k_j \delta_{i\ell}+k_i\delta_{j\ell}}{k^2}-2\frac{k_ik_jk_\ell}{k^4}\right).
\end{align}
Applying the inverse Fourier transformation in Eqs.~\eqref{theta_ft},  \eqref{p_ft} and \eqref{q_ft}--(\ref{sigma_ft}) with the help of the formulas derived in Ref.~\cite{HRG2023}, the field variables turn out to be
\begin{align}
\label{v1}
\begin{rcases}
v_i&=\frac{f_j}{\mathrm{Kn}}\left(\frac{x_i x_j}{4\pi r^2}-\frac{2\ln{r}-1}{8\pi}\delta_{ij}\right)+\frac{c_p \mathrm{Kn}}{\mathrm{Pr}} \alpha_0^2 \frac{f_j}{2\pi} \left(\frac{2x_i x_j}{r^4}-\frac{\delta_{ij}}{r^2}\right),
\\[1ex]
q_i&=-\frac{c_p \mathrm{Kn}}{\mathrm{Pr}} \alpha_0 \frac{f_j}{2\pi}\left(\frac{2x_i x_j}{r^4}-\frac{\delta_{ij}}{r^2}\right),
\\[1ex]
p&=\frac{f_i x_i}{2\pi r^2},
\\[1ex]
T&=0,
\\[1ex]
\sigma_{ij}&=\frac{f_\ell x_\ell}{2\pi}\left(\frac{2x_ix_j}{r^4}-\frac{\delta_{ij}}{r^2}\right),
\end{rcases}\text{Case I}
\end{align}
where $r=|x_i|$ and $i,j,\ell \in \{1,2\}$.
In the second case, a sourcing term is considered in the energy balance equation; so the balance equations in this case read
\begin{align}
\label{mass_bl2}
\frac{\partial v_i}{\partial x_i}&=0,
\\
\label{mom_bl2}
\frac{\partial p}{\partial x_i}+\frac{\partial \sigma_{ij}}{\partial x_j}&=0,
\\
\label{energy_bl2}
\frac{\partial q_i}{\partial x_i}&=g \,\delta(\bm{r}).
\end{align}
Applying the Fourier transformation in Eqs.~\eqref{mass_bl2}--\eqref{energy_bl2}, \eqref{ccr_1_q2D} and \eqref{ccr_2_q2D} in this case,  we obtain 
\begin{align}
\label{mass_ft2}
k_i \hat{v}_i &=0,
\\
\label{mom_ft2}
k_i \hat{p} + k_j \hat{\sigma}_{ij} &= 0,
\\
\label{energy_ft2}
k_i \hat{q}_i &= \mathbbm{i} \, g.
\end{align}
Using Eqs.~\eqref{mass_ft2} and \eqref{energy_ft2}, Eq.~\eqref{ccr1_ft} simplifies to
\begin{align}
\label{ss1_2}
\hat{\sigma}_{ij}=& \medspace\mathbbm{i} \,\mathrm{Kn} \big[ k_j(\hat{v}_i + \alpha_0\hat{q}_i)+k_i(\hat{v}_j+\alpha_0\hat{q}_j)\big]+\frac{2}{3} \delta_{ij} \mathrm{Kn}\alpha_0 g.
\end{align}
Multiplying the above equation with  $k_j$ and $k_i k_j$, we obtain \begin{align}
\label{ss11_2}
k_j\hat{\sigma}_{ij}&=\mathbbm{i} \, \mathrm{Kn}\, k^2(\hat{v}_i+\alpha_0\hat{q}_i) - \frac{1}{3}\mathrm{Kn}\, k_i\alpha_0 g,
\\
\label{ss12_2}
k_i k_j\hat{\sigma}_{ij}&=-\frac{4}{3}\mathrm{Kn}\, k^2 \alpha_0 g.
\end{align}
Multiplying Eq.~\eqref{ccr2_ft} with $k_i$ and exploiting Eqs.~\eqref{energy_ft2} and \eqref{ss12_2},  we obtain 
\begin{align}
\label{theta_ft2}
\hat{T}=\frac{\mathrm{Pr}}{c_p \mathrm{Kn}} \frac{g}{k^2} + \frac{4}{3} \mathrm{Kn}\alpha_0^2 g.
\end{align}
Again, multiplying Eq.~\eqref{mom_ft} with $k_i$ and exploiting Eq.~\eqref{ss12_2},  we obtain 
\begin{align}
\label{p_ft2}
\hat{p}&= \frac{4}{3}\mathrm{Kn}\alpha_0 g.
\end{align}
Now, from Eqs.~\eqref{mom_ft} and \eqref{p_ft2}, one can easily write \begin{align}
\label{ss5_2}
k_j\hat{\sigma}_{ij}&=-\frac{4}{3}k_i \mathrm{Kn}\alpha_0 g.
\end{align}
Substituting the value of $\hat{T}$ from Eq.~\eqref{theta_ft2} and the value of $k_j\hat{\sigma}_{ij}$ from Eq.~\eqref{ss5_2} into Eq.~\eqref{ccr2_ft}, we obtain
\begin{align}
\label{q_ft2}
\hat{q}_i&=\mathbbm{i}\frac{k_i g}{k^2} .
\end{align}
Now, from Eqs.~\eqref{ss11_2}, \eqref{ss5_2} and \eqref{q_ft2},
\begin{align}
\label{v_ft2}
\hat{v}_i=& 0.
\end{align}
Finally,  using Eqs.~\eqref{q_ft2} and \eqref{v_ft2} in Eq.~\eqref{ccr1_ft}, we obtain \begin{align}
\label{sigma_ft2}
\hat{\sigma}_{ij}=-2\mathrm{Kn}\left(\frac{k_ik_j}{k^2}-\frac{\delta_{ij}}{3}\right)\alpha_0 g.
\end{align}%
Applying the inverse Fourier transformation in Eqs.~\eqref{theta_ft2},  \eqref{p_ft2} and \eqref{q_ft2}--(\ref{sigma_ft2}) with the help of the formulae derived in~\citep{HRG2023}, the field variables turn out to be
\begin{align}
\begin{rcases}
\label{v2}
v_i&=0,
\\[1ex]
q_i&=\frac{g}{2\pi} \frac{x_i}{r^2},
\\[1ex]
p&=0,
\\[1ex]
T&=-\frac{\mathrm{Pr}}{c_p\mathrm{Kn}} \frac{g\,\ln{r}}{2\pi},
\\[1ex]
\sigma_{ij}&=\frac{2\mathrm{Kn}\alpha_0 g}{2\pi}\left(\frac{2x_ix_j}{r^4}-\frac{\delta_{ij}}{r^2}\right).
\end{rcases}\text{Case II}
\end{align}
Combining the two cases~\eqref{v1} and~\eqref{v2}, we obtain the fundamental solutions~\eqref{vel}--\eqref{heat}.
{\color{black}
\section{\label{App2}Sensitivity of the results towards the  location of singularities}
Placement of singularity points is a crucial factor in achieving accurate results through the MFS~\citep{CKL2016, WLQ2018, CH2020} due to the fact that the linear system generated on applying the MFS is often accompanied with an ill-conditioned coefficient matrix.
Thus, there is a trade-off between the well conditioning of the matrix and the accuracy of the results obtained from the MFS quite often~\citep{Alves2009}.
This trade-off was also examined in our previous work~\citep{HRG2023}. 
To analyze the sensitivity of the results obtained through the MFS to the location of singularity points as well as to the numbers of boundary nodes and singularities in the present work, we perform an error analysis for the problem of flow past a circular cylinder.

The MFS can yield reliable results even when the collocation matrix $\mathcal{M}$ in system~\eqref{mat_eq} has a high condition number, thus the condition number alone is not a sufficient indicator of accuracy. 
A more appropriate measure of the accuracy is the effective condition number instead.
The concept of the effective condition number in the context of identifying an optimal location for placement of singularity points has also been explored in the literature (see, e.g., Refs.~\cite{WL2011, CNYC2023})---hypothesizing an inverse relationship between the inaccuracy of the MFS and the effective condition number.
Following the same technique, we investigate the relationship between the effective condition number and absolute error in the speed for the case of flow past circular cylinder---aiming to get an appropriate location for the placement of singularity points.

The effective condition number for system~\eqref{mat_eq} is determined by using the singular value decomposition of the matrix $\mathcal{M}$ and the vector $\bm{b}$ appearing on the right-hand side of system~\eqref{mat_eq}. 
Let the singular value decomposition of the matrix $\mathcal{M}$ be $\mathcal{M} = \mathcal{A} \mathcal{D} \mathcal{B}^\mathsf{T}$, where $\mathcal{A}$ and $\mathcal{B}$ are the orthogonal matrices of dimensions $3N_b\times 3N_b$ and $3N_s\times 3N_s$, respectively, and $\mathcal{D}$ is a rectangular diagonal matrix of dimension $3N_b\times 3N_s$ containing the positive singular values in descending order: $\sigma_1\geq \sigma_2\geq\sigma_3\geq\dots \geq \sigma_r>0$ with $r$ being less than or equal to the minimum of $3N_b$ and $3N_s$ (i.e.~$r \leq 3\min\{N_b,N_s\}$).
Then the effective condition number for system~\eqref{mat_eq} is given by $\kappa_{\mathrm{eff}} = \Vert\bm{b}\Vert/(\sigma_r \Vert\bm{U}\Vert)$. 
We also define the absolute error $\epsilon$ in the speed of the flow by $\epsilon=|\text{speed}_{\mathrm{MFS}}-\text{speed}_{\mathrm{analytic}}|$ and introduce a dilation parameter $\alpha>1$, which relates the radii of the inner and outer fictitious boundaries (containing singularities) to the radii of the inner and outer actual boundaries via the relations $R_1'=R_1/\alpha$ and $R_2'=\alpha R_2$.
%
In what follows, we examine the changes in the effective condition number $\kappa_{\mathrm{eff}}$ and in the absolute error $\epsilon$ on changing the dilation parameter for different numbers of the boundary and source points in the cases when the total number of boundary points are the same as the total number of the singularity points (i.e.,~$N_b=N_s$) and when the total number of boundary points are different from the total number of the singularity points (i.e.,~$N_b \neq N_s$).


Figure~\ref{fig:eff_cond_error1} illustrates the effective condition number (left panel) and absolute error in the speed (right panel) both plotted against the dilation parameter in the first case when the total number of boundary points is the same as the total number of the singularity points (i.e.,~$N_b=N_s$) for different values of the number of boundary nodes (or singularity points)---specifically, for $N_b=N_s=120$, $150$ and $180$.
The left panel of the figure shows that, for small $\alpha$ (close to $1$), the effective condition number is relatively small (with $\mathcal{O}(10^5)$), and it increases rapidly with increasing the dilation parameter but peaks for $\alpha$ values somewhere in between $\alpha \approx 1.5$ and $\alpha \approx 2$ for all the considered numbers of boundary (or singularity) points, attaining maximum values of $\mathcal{O}(10^{12})$.
For $\alpha \gtrsim 2$, the effective condition number tends to stabilize a bit and starts to decrease slightly with increasing the dilation parameter.
The right panel of Fig.~\ref{fig:eff_cond_error1} shows that, for small $\alpha$ (close to $1$), the absolute error in the speed is relatively large (with $\mathcal{O}(10^{-3})$), and it decreases sharply with increasing the dilation parameter but bottoms out for $\alpha$ values somewhere in between $\alpha \approx 1.9$ and $\alpha \approx 2.7$  for all the considered numbers of boundary (or singularity) points, attaining values of $\mathcal{O}(10^{-15})$.
For even larger values of the dilation parameter, the error remains at $\mathcal{O}(10^{-15})$ and no significant improvement in the accuracy is achieved on increasing the dilation parameter further.
\begin{figure}[!tb]
\centering
\includegraphics[height=55mm]{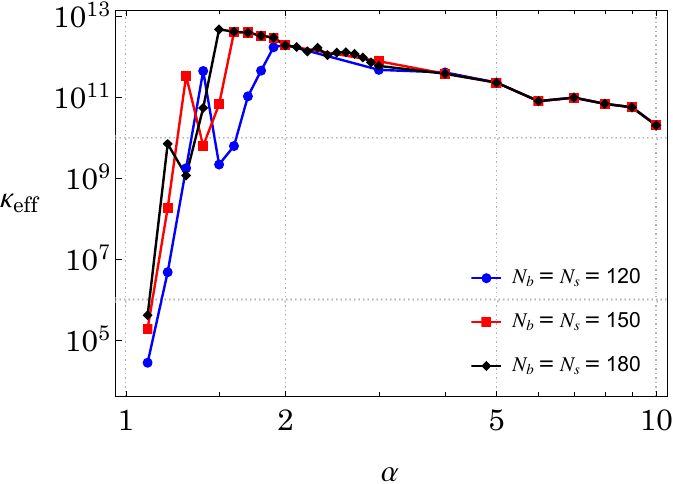}
\hfill
\includegraphics[height=55mm]{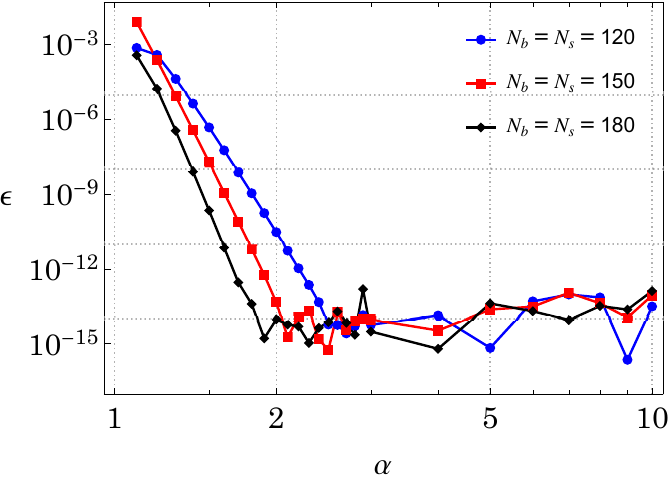}
{\color{black}\caption{\label{fig:eff_cond_error1} 
Effective condition number $\kappa_{\mathrm{eff}}$ (left panel) and the absolute error $\epsilon$ in speed (right panel) both plotted against the dilation parameter $\alpha$ in the case when the total number of boundary nodes $N_b$ is equal to the total number of singularity points $N_s$ (the case of square collocation matrix) for $\mathrm{Kn}=0.1$.
%
}}
\end{figure}



\begin{figure}[!tb]
\centering
\includegraphics[height=55mm]{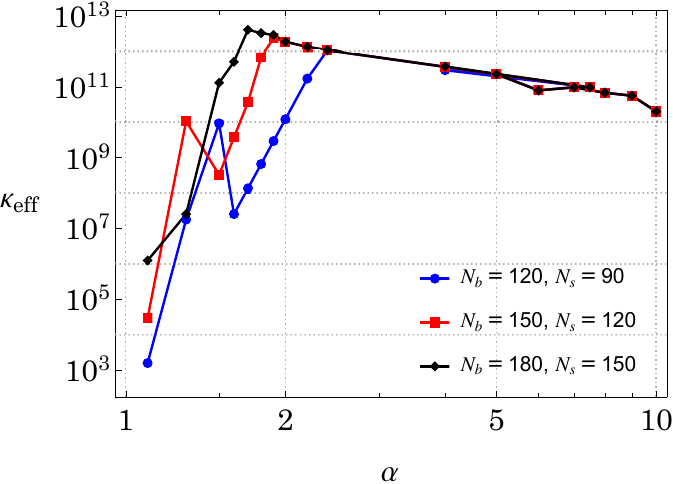}\hfill
\includegraphics[height=55mm]{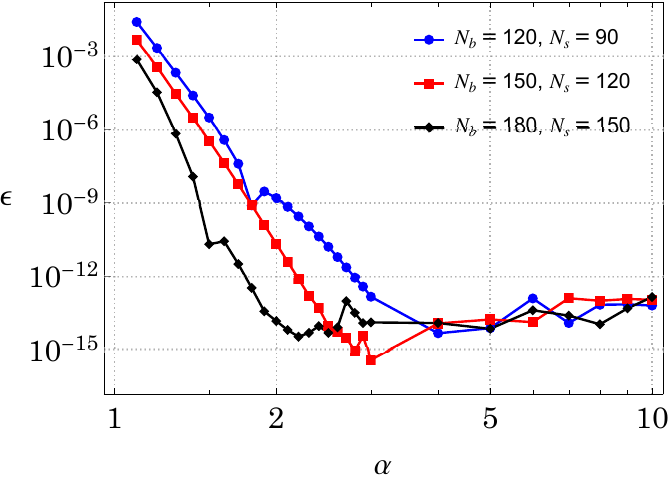}
{\color{black}\caption{\label{fig:eff_cond_error2} 
Effective condition number $\kappa_{\mathrm{eff}}$ (left panel) and the absolute error $\epsilon$ in speed (right panel) both plotted against the dilation parameter $\alpha$ in the case when the total number of boundary nodes $N_b$ is different from the total number of singularity points $N_s$ (the case of non-square collocation matrix) for $\mathrm{Kn}=0.1$.
}}
\end{figure}
Figure~\ref{fig:eff_cond_error2} also displays the effective condition number (left panel) and absolute error in the speed (right panel) both plotted against the dilation parameter but in the second case when the total number of boundary points is different from the total number of the singularity points (i.e.,~$N_b \neq N_s$). 
We have considered three combinations of the numbers of boundary nodes and singularity points, namely (i)~$N_b=120$ and $N_s=90$, (ii)~$N_b=150$ and $N_s=120$, and (iii)~$N_b=180$ and $N_s=150$. 
Similarly to the left panel in Fig.~\ref{fig:eff_cond_error1}, the effective condition number is relatively small (with $\mathcal{O}(10^3)$-$\mathcal{O}(10^6)$) for the dilation parameter close to $1$, and increases sharply with increasing $\alpha$ but peaks for $\alpha$ values somewhere in between $\alpha \approx 1.7$ and $\alpha \approx 2.2$ for all the considered numbers of boundary and singularity points, attaining maximum values of $\mathcal{O}(10^{12})$.
Similarly to the right panel in Fig.~\ref{fig:eff_cond_error1}, the absolute error in the speed is relatively large (with $\mathcal{O}(10^{-3})$) for the dilation parameter close to $1$, and it decreases sharply with increasing the dilation parameter but bottoms out for $\alpha$ values somewhere in between $\alpha \approx 2.1$ and $\alpha \approx 3$  for all the considered combinations of the numbers of boundary and singularity points, attaining values of $\mathcal{O}(10^{-15})$.
For even larger values of the dilation parameter, the error remains at $\mathcal{O}(10^{-15})$ with a slightly increasing trend and hence no significant improvement in the accuracy is achieved on increasing the dilation parameter further.


Both Figs.~\ref{fig:eff_cond_error1} and \ref{fig:eff_cond_error2} exhibit an inverse relationship between the effective condition number and absolute error, which is concurrent with the findings of \cite{WL2011, CNYC2023}.
Noticing the trends in the effective condition number and absolute error, we choose $N_b=N_s=150$ as it ensures a sufficiently high number of boundary nodes and singularities in order to achieve a high effective condition number and better accuracy along with computational efficiency.
Additionally, the choice of $\alpha = 2$ balances the trade-off between achieving a high condition number and minimizing the error.


 }

\bibliographystyle{rspublicnatwithsort_implicitdoi}
\bibliography{references}

\begin{thebibliography}{47}
\providecommand{\natexlab}[1]{#1}
\expandafter\ifx\csname urlstyle\endcsname\relax
  \providecommand{\doi}[1]{doi:\discretionary{}{}{}#1}\else
  \providecommand{\doi}{doi:\discretionary{}{}{}\begingroup
  \urlstyle{rm}\Url}\fi

\bibitem[{Stokes(1851)}]{stokes1851}
Stokes, G.~G. 1851 On the effect of the internal friction of fluids on the
  motion of pendulums.
\newblock \emph{Trans. Cambridge Philos. Soc.}, \textbf{9 (Part II)}, 8--106.

\bibitem[{Tanner(1993)}]{Tanner1993}
Tanner, R. 1993 Stokes paradox for power-law flow around a cylinder.
\newblock
  \href{http://dx.doi.org/10.1016/0377-0257(93)80032-7}{\emph{\href{http://dx.doi.org/10.1016/0377-0257(93)80032-7}{J.
  Nonnewton. Fluid Mech.}}},
  \href{http://dx.doi.org/10.1016/0377-0257(93)80032-7}{\textbf{50}, 217--224}.

\bibitem[{Van~Dyke(1964)}]{VanDyke1964}
Van~Dyke, M. 1964 \emph{Perturbation Methods in Fluid Mechanics}.
\newblock New York: Academic Press.

\bibitem[{Lamb(1932)}]{Lamb1932}
Lamb, H. 1932 \emph{Hydrodynamics}.
\newblock Cambridge: Cambridge University Press.

\bibitem[{Maru\v{s}i{\'c}-Paloka(2001)}]{Paloka2001}
Maru\v{s}i{\'c}-Paloka, E. 2001 On the {S}tokes paradox for power-law fluids.
\newblock
  \href{http://dx.doi.org/https://doi.org/10.1002/1521-4001(200101)81:1<31::AID-ZAMM31>3.0.CO;2-G}{\emph{\href{http://dx.doi.org/https://doi.org/10.1002/1521-4001(200101)81:1<31::AID-ZAMM31>3.0.CO;2-G}{Z.
  Angew. Math. Mech.}}},
  \href{http://dx.doi.org/https://doi.org/10.1002/1521-4001(200101)81:1<31::AID-ZAMM31>3.0.CO;2-G}{\textbf{81},
  31--36}.

\bibitem[{Smith(1990)}]{Smith1990}
Smith, S.~H. 1990 Some limitations of two-dimensional unbounded {S}tokes flow.
\newblock
  \href{http://dx.doi.org/10.1063/1.857699}{\emph{\href{http://dx.doi.org/10.1063/1.857699}{Phys.
  Fluids A}}}, \href{http://dx.doi.org/10.1063/1.857699}{\textbf{2},
  1724--1730}.

\bibitem[{Morra(2018)}]{Morra2018}
Morra, G. 2018 Insights on the physics of {S}tokes flow.
\newblock In \emph{Pythonic Geodynamics: Implementations for Fast Computing},
  pp. 93--104. Cham: Springer International Publishing.
\newblock \doi{10.1007/978-3-319-55682-6_6}.

\bibitem[{Oseen(1910)}]{oseen1910}
Oseen, C.~W. 1910 Uber die {S}tokes'sche {F}ormel und {\"u}ber eine verwandte
  {A}ufgabe in der {H}ydrodynamik.
\newblock \emph{Arkiv Mat. Astron. Fysk.}, \textbf{6}, 1--59.

\bibitem[{Lamb(1911)}]{lamb1911}
Lamb, H. 1911 {XV.} on the uniform motion of a sphere through a viscous fluid.
\newblock
  \href{http://dx.doi.org/10.1080/14786440108637012}{\emph{\href{http://dx.doi.org/10.1080/14786440108637012}{London,
  Edinburgh, Dublin Philos. Mag. J. Sci.}}},
  \href{http://dx.doi.org/10.1080/14786440108637012}{\textbf{21}, 112--121}.

\bibitem[{Bairstow \emph{et~al.}(1923)Bairstow, Cave \& Lang}]{BCL1923}
Bairstow, L., Cave, B.~M. \& Lang, E.~D. 1923 X. {T}he resistance of a cylinder
  moving in a viscous fluid.
\newblock
  \href{http://dx.doi.org/10.1098/rsta.1923.0010}{\emph{\href{http://dx.doi.org/10.1098/rsta.1923.0010}{Philos.
  Trans. Royal Soc. A}}},
  \href{http://dx.doi.org/10.1098/rsta.1923.0010}{\textbf{223}, 383--432}.

\bibitem[{Tomotika \& Aoi(1950)}]{TA1950}
Tomotika, S. \& Aoi, T. 1950 The steady flow of viscous fluid past a sphere and
  circular cylinder at small {R}eynolds numbers.
\newblock
  \href{http://dx.doi.org/10.1093/qjmam/3.2.141}{\emph{\href{http://dx.doi.org/10.1093/qjmam/3.2.141}{Q.
  J. Mech. Appl. Math.}}},
  \href{http://dx.doi.org/10.1093/qjmam/3.2.141}{\textbf{3}, 141--161}.

\bibitem[{Kaplun(1954)}]{Kaplun1954}
Kaplun, S. 1954 The role of coordinate systems in boundary-layer theory.
\newblock
  \href{http://dx.doi.org/10.1007/BF01600771}{\emph{\href{http://dx.doi.org/10.1007/BF01600771}{Z.
  Angew. Math. Phys.}}},
  \href{http://dx.doi.org/10.1007/BF01600771}{\textbf{5}, 111--135}.

\bibitem[{Kaplun \& Lagerstrom(1957)}]{KL1957}
Kaplun, S. \& Lagerstrom, P.~A. 1957 Asymptotic expansions of {N}avier-{S}tokes
  solutions for small {R}eynolds numbers.
\newblock
  \href{http://dx.doi.org/10.1512/IUMJ.1957.6.56028}{\emph{\href{http://dx.doi.org/10.1512/IUMJ.1957.6.56028}{Appl.
  Math. Mech.}}},
  \href{http://dx.doi.org/10.1512/IUMJ.1957.6.56028}{\textbf{6}, 585--593}.

\bibitem[{Proudman \& Pearson(1957)}]{PP1957}
Proudman, I. \& Pearson, J. R.~A. 1957 Expansions at small {R}eynolds numbers
  for the flow past a sphere and a circular cylinder.
\newblock
  \href{http://dx.doi.org/10.1017/S0022112057000105}{\emph{\href{http://dx.doi.org/10.1017/S0022112057000105}{J.
  Fluid Mech.}}},
  \href{http://dx.doi.org/10.1017/S0022112057000105}{\textbf{2}, 237--262}.

\bibitem[{Kida \& Take(1992)}]{KT1992}
Kida, T. \& Take, T. 1992 Asymptotic expansions for low {R}eynolds number flow
  past a cylindrical body.
\newblock
  \href{http://dx.doi.org/10.1299/jsmeb1988.35.2_144}{\emph{\href{http://dx.doi.org/10.1299/jsmeb1988.35.2_144}{JSME
  Int. J. Ser. 2}}},
  \href{http://dx.doi.org/10.1299/jsmeb1988.35.2_144}{\textbf{35}, 144--150}.

\bibitem[{Khalili \& Liu(2017)}]{KL2017}
Khalili, A. \& Liu, B. 2017 {Stokes'} paradox: {C}reeping flow past a
  two-dimensional cylinder in an infinite domain.
\newblock
  \href{http://dx.doi.org/10.1017/jfm.2017.127}{\emph{\href{http://dx.doi.org/10.1017/jfm.2017.127}{J.
  Fluid Mech.}}}, \href{http://dx.doi.org/10.1017/jfm.2017.127}{\textbf{817},
  374--387}.

\bibitem[{Cercignani(1968)}]{Cercignani1968}
Cercignani, C. 1968 Stokes paradox in kinetic theory.
\newblock
  \href{http://dx.doi.org/10.1063/1.1691903}{\emph{\href{http://dx.doi.org/10.1063/1.1691903}{Phys.
  Fluids}}}, \href{http://dx.doi.org/10.1063/1.1691903}{\textbf{11}, 303--308}.

\bibitem[{Yamamoto \& Sera(1985)}]{YS1985}
Yamamoto, K. \& Sera, K. 1985 Flow of a rarefied gas past a circular cylinder.
\newblock
  \href{http://dx.doi.org/10.1063/1.865012}{\emph{\href{http://dx.doi.org/10.1063/1.865012}{Phys.
  Fluids}}}, \href{http://dx.doi.org/10.1063/1.865012}{\textbf{28},
  1286--1293}.

\bibitem[{Gu \emph{et~al.}(2019)Gu, Barber, John \& Emerson}]{GBJE2019}
Gu, X.-J., Barber, R.~W., John, B. \& Emerson, D.~R. 2019 Non-equilibrium
  effects on flow past a circular cylinder in the slip and early transition
  regime.
\newblock
  \href{http://dx.doi.org/10.1017/jfm.2018.869}{\emph{\href{http://dx.doi.org/10.1017/jfm.2018.869}{J.
  Fluid Mech.}}}, \href{http://dx.doi.org/10.1017/jfm.2018.869}{\textbf{860},
  654--681}.

\bibitem[{Rana \emph{et~al.}(2018)Rana, Gupta \& Struchtrup}]{RGS2018}
Rana, A.~S., Gupta, V.~K. \& Struchtrup, H. 2018 Coupled constitutive
  relations: a second law based higher-order closure for hydrodynamics.
\newblock
  \href{http://dx.doi.org/10.1098/rspa.2018.0323}{\emph{\href{http://dx.doi.org/10.1098/rspa.2018.0323}{Proc.
  Roy. Soc. A}}}, \href{http://dx.doi.org/10.1098/rspa.2018.0323}{\textbf{474},
  20180\,323}.

\bibitem[{Kupradze \& Aleksidze(1964)}]{KA1964}
Kupradze, V.~D. \& Aleksidze, M.~A. 1964 The method of functional equations for
  the approximate solution of certain boundary value problems.
\newblock
  \href{http://dx.doi.org/10.1016/0041-5553(64)90006-0}{\emph{\href{http://dx.doi.org/10.1016/0041-5553(64)90006-0}{USSR
  Comput. Math. Math. Phys.}}},
  \href{http://dx.doi.org/10.1016/0041-5553(64)90006-0}{\textbf{4}, 82--126}.

\bibitem[{Berger \& Karageorghis(2001)}]{BK2001}
Berger, J.~R. \& Karageorghis, A. 2001 The method of fundamental solutions for
  layered elastic materials.
\newblock
  \href{http://dx.doi.org/10.1016/S0955-7997(01)00002-9}{\emph{\href{http://dx.doi.org/10.1016/S0955-7997(01)00002-9}{Eng.
  Anal. Bound. Elem.}}},
  \href{http://dx.doi.org/10.1016/S0955-7997(01)00002-9}{\textbf{25},
  877--886}.

\bibitem[{Fairweather \emph{et~al.}(2003)Fairweather, Karageorghis \&
  Martin}]{FKM2003}
Fairweather, G., Karageorghis, A. \& Martin, P. 2003 The method of fundamental
  solutions for scattering and radiation problems.
\newblock
  \href{http://dx.doi.org/10.1016/S0955-7997(03)00017-1}{\emph{\href{http://dx.doi.org/10.1016/S0955-7997(03)00017-1}{Eng.
  Anal. Bound. Elem.}}},
  \href{http://dx.doi.org/10.1016/S0955-7997(03)00017-1}{\textbf{27},
  759--769}.

\bibitem[{Karageorghis \emph{et~al.}(2011)Karageorghis, Lesnic \&
  Marin}]{KLM2011}
Karageorghis, A., Lesnic, D. \& Marin, L. 2011 A survey of applications of the
  {MFS} to inverse problems.
\newblock
  \href{http://dx.doi.org/10.1080/17415977.2011.551830}{\emph{\href{http://dx.doi.org/10.1080/17415977.2011.551830}{Inverse
  Probl. Sci. Eng.}}},
  \href{http://dx.doi.org/10.1080/17415977.2011.551830}{\textbf{19}, 309--336}.

\bibitem[{Liu \emph{et~al.}(2021)Liu, Fan \& {\v S}arler}]{LFS2021}
Liu, Q.~G., Fan, C.~M. \& {\v S}arler, B. 2021 Localized method of fundamental
  solutions for two-dimensional anisotropic elasticity problems.
\newblock
  \href{http://dx.doi.org/10.1016/j.enganabound.2021.01.008}{\emph{\href{http://dx.doi.org/10.1016/j.enganabound.2021.01.008}{Eng.
  Anal. Bound. Elem.}}},
  \href{http://dx.doi.org/10.1016/j.enganabound.2021.01.008}{\textbf{125},
  59--65}.

\bibitem[{Lockerby \& Collyer(2016)}]{LC2016}
Lockerby, D.~A. \& Collyer, B. 2016 Fundamental solutions to moment equations
  for the simulation of microscale gas flows.
\newblock
  \href{http://dx.doi.org/10.1017/jfm.2016.606}{\emph{\href{http://dx.doi.org/10.1017/jfm.2016.606}{J.
  Fluid Mech.}}}, \href{http://dx.doi.org/10.1017/jfm.2016.606}{\textbf{806},
  413--436}.

\bibitem[{Claydon \emph{et~al.}(2017)Claydon, Shrestha, Rana, Sprittles \&
  Lockerby}]{CSRSL2017}
Claydon, R., Shrestha, A., Rana, A.~S., Sprittles, J.~E. \& Lockerby, D.~A.
  2017 Fundamental solutions to the regularised 13-moment equations: efficient
  computation of three-dimensional kinetic effects.
\newblock
  \href{http://dx.doi.org/10.1017/jfm.2017.763}{\emph{\href{http://dx.doi.org/10.1017/jfm.2017.763}{J.
  Fluid Mech.}}}, \href{http://dx.doi.org/10.1017/jfm.2017.763}{\textbf{833},
  R4}.

\bibitem[{Rana \emph{et~al.}(2021{\natexlab{\emph{a}}})Rana, Saini,
  Chakraborty, Lockerby \& Sprittles}]{RSCLS2021}
Rana, A.~S., Saini, S., Chakraborty, S., Lockerby, D.~A. \& Sprittles, J.~E.
  2021{\natexlab{\emph{a}}} Efficient simulation of non-classical
  liquid--vapour phase-transition flows: a method of fundamental solutions.
\newblock
  \href{http://dx.doi.org/10.1017/jfm.2021.405}{\emph{\href{http://dx.doi.org/10.1017/jfm.2021.405}{J.
  Fluid Mech.}}}, \href{http://dx.doi.org/10.1017/jfm.2021.405}{\textbf{919},
  A35}.

\bibitem[{Himanshi \emph{et~al.}(2023)Himanshi, Rana \& Gupta}]{HRG2023}
Himanshi, Rana, A.~S. \& Gupta, V.~K. 2023 Fundamental solutions of an extended
  hydrodynamic model in two dimensions: {D}erivation, theory, and applications.
\newblock
  \href{http://dx.doi.org/10.1103/PhysRevE.108.015306}{\emph{\href{http://dx.doi.org/10.1103/PhysRevE.108.015306}{Phys.
  Rev. E}}}, \href{http://dx.doi.org/10.1103/PhysRevE.108.015306}{\textbf{108},
  015\,306}.

\bibitem[{Torrilhon(2010)}]{Torrilhon2010}
Torrilhon, M. 2010 Slow gas microflow past a sphere: {A}nalytical solution
  based on moment equations.
\newblock
  \href{http://dx.doi.org/10.1063/1.3453707}{\emph{\href{http://dx.doi.org/10.1063/1.3453707}{Phys.
  Fluids}}}, \href{http://dx.doi.org/10.1063/1.3453707}{\textbf{22}, 072\,001}.

\bibitem[{Westerkamp \& Torrilhon(2012)}]{WT2012}
Westerkamp, A. \& Torrilhon, M. 2012 Slow rarefied gas flow past a cylinder:
  {A}nalytical solution in comparison to the sphere.
\newblock
  \href{http://dx.doi.org/10.1063/1.4769505}{\emph{\href{http://dx.doi.org/10.1063/1.4769505}{AIP
  Conf. Proc.}}}, \href{http://dx.doi.org/10.1063/1.4769505}{\textbf{1501},
  207--214}.

\bibitem[{Gupta(2020)}]{Gupta2020}
Gupta, V.~K. 2020 Moment theories for a $d$-dimensional dilute granular gas of
  {M}axwell molecules.
\newblock
  \href{http://dx.doi.org/10.1017/jfm.2020.20}{\emph{\href{http://dx.doi.org/10.1017/jfm.2020.20}{J.
  Fluid Mech.}}}, \href{http://dx.doi.org/10.1017/jfm.2020.20}{\textbf{888},
  A12}.

\bibitem[{Maxwell(1879)}]{Maxwell1879}
Maxwell, J.~C. 1879 On stresses in rarefied gases arising from inequalities of
  temperature.
\newblock
  \href{http://dx.doi.org/10.1098/rstl.1879.0067}{\emph{\href{http://dx.doi.org/10.1098/rstl.1879.0067}{Phil.
  Trans. R. Soc. Lond.}}},
  \href{http://dx.doi.org/10.1098/rstl.1879.0067}{\textbf{170}, 231--256}.

\bibitem[{Sone(2007)}]{Sone2007}
Sone, Y. 2007 \emph{Molecular Gas Dynamics: Theory, Techniques and
  Applications}.
\newblock Boston: Birkh{\"a}user.

\bibitem[{Rana \emph{et~al.}(2021{\natexlab{\emph{b}}})Rana, Gupta, Sprittles
  \& Torrilhon}]{RGST2021}
Rana, A.~S., Gupta, V.~K., Sprittles, J.~E. \& Torrilhon, M.
  2021{\natexlab{\emph{b}}} {$H$}-theorem and boundary conditions for the
  linear {R26} equations: application to flow past an evaporating droplet.
\newblock
  \href{http://dx.doi.org/10.1017/jfm.2021.622}{\emph{\href{http://dx.doi.org/10.1017/jfm.2021.622}{J.
  Fluid Mech.}}}, \href{http://dx.doi.org/10.1017/jfm.2021.622}{\textbf{924},
  A16}.

\bibitem[{Cheng \& Hong(2020)}]{CH2020}
Cheng, A.~H. \& Hong, Y. 2020 An overview of the method of fundamental
  solutions---{Solvability}, uniqueness, convergence, and stability.
\newblock
  \href{http://dx.doi.org/10.1016/j.enganabound.2020.08.013}{\emph{\href{http://dx.doi.org/10.1016/j.enganabound.2020.08.013}{Eng.
  Anal. Bound. Elem.}}},
  \href{http://dx.doi.org/10.1016/j.enganabound.2020.08.013}{\textbf{120},
  118--152}.

\bibitem[{Bakanov \emph{et~al.}(1983)Bakanov, Vysotskij, Derjaguin \&
  Roldughin}]{BVDR1983}
Bakanov, S.~P., Vysotskij, V.~V., Derjaguin, B.~V. \& Roldughin, V.~I. 1983
  Thermal polarization of bodies in the rarefied gas flow.
\newblock
  \href{http://dx.doi.org/doi:10.1515/jnet.1983.8.1.75}{\emph{\href{http://dx.doi.org/doi:10.1515/jnet.1983.8.1.75}{J.
  Non-Equilib. Thermodyn.}}},
  \href{http://dx.doi.org/doi:10.1515/jnet.1983.8.1.75}{\textbf{8}, 75--83}.

\bibitem[{Rana \emph{et~al.}(2013)Rana, Torrilhon \& Struchtrup}]{RTS2013}
Rana, A., Torrilhon, M. \& Struchtrup, H. 2013 A robust numerical method for
  the {R13} equations of rarefied gas dynamics: {A}pplication to lid driven
  cavity.
\newblock
  \href{http://dx.doi.org/10.1016/j.jcp.2012.11.023}{\emph{\href{http://dx.doi.org/10.1016/j.jcp.2012.11.023}{J.
  Comput. Phys.}}},
  \href{http://dx.doi.org/10.1016/j.jcp.2012.11.023}{\textbf{236}, 169--186}.

\bibitem[{Gupta(2015)}]{Gupta2015}
Gupta, V.~K. 2015 Mathematical modeling of rarefied gas mixtures.
\newblock Ph.D. thesis, RWTH Aachen University, Germany.

\bibitem[{Pich(1969)}]{Pich1969}
Pich, J. 1969 The drag of a cylinder in the transition region.
\newblock
  \href{http://dx.doi.org/https://doi.org/10.1016/0021-9797(69)90350-6}{\emph{\href{http://dx.doi.org/https://doi.org/10.1016/0021-9797(69)90350-6}{J.
  Colloid Interface Sci.}}},
  \href{http://dx.doi.org/https://doi.org/10.1016/0021-9797(69)90350-6}{\textbf{29},
  91--96}.

\bibitem[{Fam \& Rashed(2002)}]{FR2002}
Fam, G. S.~A. \& Rashed, Y.~F. 2002 A study on the source points locations in
  the method of fundamental solution.
\newblock pp. 297--312. WIT Press, Southampton, UK.
\newblock \doi{10.2495/BE020281}.

\bibitem[{Nie \& Lin(2019)}]{NL2020}
Nie, D. \& Lin, J. 2019 Numerical study on flow past a confined half cylinder
  in two bluff arrangements.
\newblock
  \href{http://dx.doi.org/10.1088/1873-7005/ab5842}{\emph{\href{http://dx.doi.org/10.1088/1873-7005/ab5842}{Fluid
  Dyn. Res.}}}, \href{http://dx.doi.org/10.1088/1873-7005/ab5842}{\textbf{52},
  015\,506}.

\bibitem[{Chen \emph{et~al.}(2016)Chen, Karageorghis \& Li}]{CKL2016}
Chen, C.~S., Karageorghis, A. \& Li, Y. 2016 On choosing the location of the
  sources in the {MFS}.
\newblock
  \href{http://dx.doi.org/10.1007/s11075-015-0036-0}{\emph{\href{http://dx.doi.org/10.1007/s11075-015-0036-0}{Numer.
  Algorithms}}},
  \href{http://dx.doi.org/10.1007/s11075-015-0036-0}{\textbf{72}, 107--130}.

\bibitem[{Wang \emph{et~al.}(2018)Wang, Liu \& Qu}]{WLQ2018}
Wang, F., Liu, C.-S. \& Qu, W. 2018 Optimal sources in the {MFS} by minimizing
  a new merit function: Energy gap functional.
\newblock
  \href{http://dx.doi.org/10.1016/j.aml.2018.07.002}{\emph{\href{http://dx.doi.org/10.1016/j.aml.2018.07.002}{Appl.
  Math. Lett.}}},
  \href{http://dx.doi.org/10.1016/j.aml.2018.07.002}{\textbf{86}, 229--235}.

\bibitem[{Alves(2009)}]{Alves2009}
Alves, C.~J. 2009 On the choice of source points in the method of fundamental
  solutions.
\newblock
  \href{http://dx.doi.org/10.1016/j.enganabound.2009.05.007}{\emph{\href{http://dx.doi.org/10.1016/j.enganabound.2009.05.007}{Eng.
  Anal. Bound. Elem.}}},
  \href{http://dx.doi.org/10.1016/j.enganabound.2009.05.007}{\textbf{33},
  1348--1361}.

\bibitem[{Wong \& Ling(2011)}]{WL2011}
Wong, K.~Y. \& Ling, L. 2011 Optimality of the method of fundamental solutions.
\newblock
  \href{http://dx.doi.org/10.1016/j.enganabound.2010.06.002}{\emph{\href{http://dx.doi.org/10.1016/j.enganabound.2010.06.002}{Eng.
  Anal. Bound. Elem.}}},
  \href{http://dx.doi.org/10.1016/j.enganabound.2010.06.002}{\textbf{35},
  42--46}.

\bibitem[{Chen \emph{et~al.}(2023)Chen, Noorizadegan, Young \& Chen}]{CNYC2023}
Chen, C., Noorizadegan, A., Young, D. \& Chen, C.-S. 2023 On the determination
  of locating the source points of the {MFS} using effective condition number.
\newblock
  \href{http://dx.doi.org/10.1016/j.cam.2022.114955}{\emph{\href{http://dx.doi.org/10.1016/j.cam.2022.114955}{J.
  Comput. Appl. Math.}}},
  \href{http://dx.doi.org/10.1016/j.cam.2022.114955}{\textbf{423}, 114\,955}.

\end{thebibliography}
\end{document}